\documentstyle[11pt,fleqn]{article}
\newcommand{\N}{\mbox{$I\!\!N$}}             
\newcommand{\R}{\mbox{$I\!\!R$}}             
\newcommand{\C}{\mbox{$I\!\!\!\!C$}}         

\begin{document}


\hfill{\sl PREPRINT - UTM 547 }
\par
\bigskip
\par
\rm


\par
\bigskip
\LARGE
\noindent
{\bf Proof of  the symmetry of the off-diagonal 
heat-kernel and Hadamard's expansion coefficients in 
general $C^{\infty}$ Riemannian manifolds}
\par
\bigskip
\par
\rm
\normalsize



\large
\smallskip

\noindent {\bf Valter Moretti}

\large
\smallskip

\noindent 
Department of Mathematics, Trento University, and
Istituto Nazionale di Fisica Nucleare,\\
Gruppo Collegato di Trento,
I-38050 Povo (TN), Italy.\\
E-mail: moretti@science.unitn.it



\par
\bigskip
\par
\hfill{\sl February 1999}
\par
\medskip
\par\rm



\noindent
{\bf Abstract:}
We consider the problem of the 
symmetry of the off-diagonal heat-kernel coefficients as well as  
the coefficients 
corresponding to the short-distance-divergent part of the Hadamard expansion
in general smooth (analytic or not) manifolds.
The requirement of such a
 symmetry played a central r\^{o}le  in the theory of the point-splitting 
one-loop renormalization of the stress tensor in either Riemannian  or 
Lorentzian manifolds. Actually, the symmetry of these coefficients has been 
assumed as a hypothesis
 in several papers concerning these issues without an explicit proof. 
The difficulty of a direct proof is related to the fact that the considered 
off-diagonal heat-kernel expansion, also in the  Riemannian case, 
in principle, may be not a proper asymptotic expansion. On the other hand, 
direct computations of the off-diagonal heat-kernel coefficients are 
impossibly difficult in nontrivial cases and thus no case is known 
in the literature where the  symmetry does not hold.
By approximating $C^\infty$ metrics with analytic metrics in common (totally 
normal) geodesically convex neighborhoods, it is rigorously proven that, in 
general $C^\infty$ Riemannian manifolds, any point admits a geodesically 
convex neighborhood where the off-diagonal heat-kernel coefficients, as well 
as the relevant Hadamard's expansion coefficients, are symmetric functions of 
the two arguments. 

\par

 \rm





\section*{Introduction.}

After earlier works (e.g. see \cite{wald78}),
 the symmetry of the  coefficients
which appear in the short-distance-divergent part  of the 
Hadamard expansion of the two-point functions of a quantum state
in curved spacetime, 
has been tacitly assumed to hold in mathematical-physics literature.
This symmetry plays a central r\^{o}le  in the renormalization  
procedure of the one-loop stress tensor in curved spacetime, 
 either in Lorentzian and 
Euclidean Quantum Field Theory. In fact, it is directly related to
 the conservation 
of the stress tensor and the appearance of the conformal anomaly 
\cite{wald78}. On the other hand, the symmetry of Hadamard 
coefficients is related to 
that of the heat-kernel coefficients \cite{bd,fu,m1,m2}. 
Despite the relevance of this assumption, 
to the knowledge of the author, up to now 
no rigorous proof  of these symmetries exists in the literature\footnote{For 
instance, in \cite{wald78}, 
such a symmetry was (indirectly) argued to hold
for the analytic case. In
\cite{fsw}, the symmetry was
 argued to hold for the $C^\infty$ case. Neverthenless, these papers did not
 report the corresponding proof.
The literature concerning the point-splitting procedure
successive to \cite{wald78}, as
\cite{bo},
assumes that symmetry implicitly.}.
In this paper,
 we shall see that the problem of the symmetry of the heat-kernel/Hadamard
coefficients  is not so trivial as it seems at first
sight. That is related to the fact that, in principle,
 the heat-kernel expansion could be
 not asymptotic in the rigorous sense, 
also in the Riemannian case, when it is performed 
off-diagonal.
We shall prove that the heat-kernel coefficients, in the Riemannian case,
are actually symmetric in a geodesically convex neighborhood of any
point of a $C^\infty$  manifold. As a result we shall also see
that the  requirement of analyticity 
of the manifold assumed in earlier work \cite{wald78,bo} can be completely
dropped (as argued in \cite{fsw}).\\

\section{Generalities, the problem of the symmetry of the heat-kernel 
and Hadamard's coefficients }

\noindent {\bf 1.1} {\em Notations, general hypoteses and preliminaries.}
Within this work, ${\cal M}$  denotes a (Hausdorff, paracompact, connected,
 orientable) $D$-dimensional $C^{\infty}$ manifold endowed with a non-singular 
metric ${\bf g}$ which makes ${\cal M}$ geodesically complete.
${\cal M}$ can be  a manifold with smooth boundary $\partial 
{\cal M}$ and we shall consider ${\bf g} \equiv g_{ab}$
either  Lorentzian or  Riemannian.

We shall deal with differential operators
of the form
\begin{eqnarray}
A_0 = - \Delta + V\:\: : \:\: C_0^{\infty}({\cal M})  \rightarrow 
L^2({\cal M}, d\mu_g) \label{caseR}\:,
\end{eqnarray}
if ${\cal M}$ is Riemannian,
 and
\begin{eqnarray}
A_0 = - \Delta + V\:\: : \:\: D({\cal M})  
\rightarrow C^{\infty}({\cal M})  \label{caseL}\:,
\end{eqnarray}
if ${\cal M}$ is Lorentzian.  $D({\cal M})$ 
is any domain of smooth
functions, like $C^\infty_0({\cal M})$ or $C^\infty({\cal M})$.
$\Delta := \nabla_a\nabla^{a}$ denotes the Laplace-Beltrami operator and
 $\nabla$ means the covariant derivative associated to the metric
connection. $d \mu_{g} $ denotes the Borel measure induced by
the metric, and $V$ is a {\em real}
function of  $C^{\infty}({\cal M})$.
Whenever ${\cal M}$ is Riemannian, we suppose also that 
$A_0$ is positive, namely, that
it 
is bounded below by some constant $C\geq 0$ (for sufficient conditions
for this requirement see \cite{de}).

The requirents  above are the {\bf general hypotheses} which we shall 
refer to throughout this paper.\\

Let us  give some definitions used throughout this work 
and recall  some known and useful relevant results.

In this paper, a {\em manifold  with boundary}  
$({\cal M},\partial {\cal M})$ is defined 
by giving a pair $({\cal N}_{\cal M},f_{\cal M})$, where 
${\cal N}_{\cal M}$ is a manifold
and $f_{\cal M} :{\cal N} \to \R$ denotes a differentiable function. The set 
${\cal M}$ is defined by ${\cal M} := \{ p\in {\cal N}_{\cal M}\:|\: 
f_{\cal M}(p) \geq 0 \}$
and the boundary $\partial {\cal M}$ is defined by
$\partial{\cal M} := \{ p\in {\cal N}_{\cal M}\:|\: f_{\cal M}(p) =  0\}$.

Throughout the paper an analytic function is a real-valued function which
 admits a (multivariable) Taylor expansion in a neighborhood of any 
point of its domain. Moreover, 
 ``smooth''  means $C^\infty$ whenever we do not specify further.

In the Riemannian case, $A_0$ is symmetric and admits
self-adjoint extensions \cite{m1}. In particular, following 
the spirit of \cite{wald78} and \cite{m1,m2}, at least in the case 
$\partial {\cal M} = \emptyset$ we shall deal with
the Friedrichs extension
 \cite{rs} {\em which will be denoted by $A$} throughout this paper. 
We remind that $A$
has the same lower bound than $A_0$. Moreover $A_0$ is essentially 
self-adjoint and thus $A$ is its unique self-adjoint extension  whenever 
either ${\cal M}$  is compact \cite{m1} or  $V \equiv 0$ \cite{de}.

Concerning derivative operators, we shall employ the following notations,
 in a fixed
local coordinate system,
\begin{eqnarray}
D^{\alpha}_x := \frac{\partial^{|\alpha|} }{\partial x^{1\alpha_1} \cdots
 \partial x^{D\alpha_D} }\vert_x
\end{eqnarray}
where the {\em multi-index} $\alpha$ is defined by
 $\alpha :=(\alpha_1,\cdots,\alpha_D)$, any
 $\alpha_i \geq 0$ being a natural number ($i=1,\cdots,D)$ and
 $|\alpha| := \alpha_1+ \cdots + \alpha_D $.

Whenever $I\subset {\cal N}$  is a {\em closed} subset of the manifold
 ${\cal N}$, 
$f\in C^k(I;\R^n)$ indicates a $\R^n$-valued  function defined on $I$
which admits
a $C^k$ extension on some {\em open} set $I'\subset {\cal N}$ such that
$I\subset I'$.

Finally,   in a fixed coordinate system,
$\nabla f$ indicates the function which
maps any point $q$ to the {\em Jacobian matrix} evaluated at $q$  of the 
function $f : p\mapsto f(p)$.\\

In any   manifold ${\cal M}$ 
endowed with a (not necessarily metric) 
affine connection which makes it geodesically complete,
the notion of {\em normal neighborhood} centered on a point $p$, 
${\cal N}_p$, 
indicates any {\em open}  neighborhood 
 of the point $p\in {\cal M}$ of the form  ${\cal N}_p =
  exp_p(B)$, $B \subset T_p({\cal M})$ being an open starshaped neighborhood
of the  origin such that  $exp_p$ defines a diffeomorphism therein.
Then, the  components of the vectors $v\in T_p({\cal M})$ contained in 
$B$ define {\em normal coordinates} on ${\cal M}$ centered in  $p$ 
via the function $v\mapsto exp_p v$. 
Notice that any $q\in {\cal N}_p$
can be connected with $p$ by only one geodesic segment
completely contained in ${\cal N}_p$. It minimize the length
of the class of  curves connecting these two 
points when the connection is metric, the metric is 
 Riemannian and $B$ is a geodesical ball. 
A {\em totally normal neighborhood} of a point $p\in {\cal M}$
 is a  neighborhood\footnote{In this work, 
a neighborhood
of a point is any set which includes an open set containing 
the point.} of $p$, 
${\cal V}_p \subset {\cal M}$, 
 such that,  for any $q\in {\cal V}_p$,
there  is a normal neighborhood centered on  $q$ containing ${\cal V}_p$. 
Therefore, if $q$ and $q'$
belong to the same totally normal neighborhood, there is only one 
 geodesic segment connecting these two points completely 
contained in any normal 
neighborhood sufficiently large centered  on each of 
 the points (but this segment is not  necessarily contained
 in ${\cal V}_p$).  Notice that  a coordinate system
which covers any totally normal neighborhood does exist in any case: 
It is that defined in a 
sufficiently large normal neighborhood of one of its points.
Finally, a {\em geodesically convex} neighborhood of a point $p\in {\cal M}$
 is a totally normal neighborhood of $p$,
 ${\cal U}_p$, such that, for any pair $q,q'\in {\cal U}_p$, 
there is only one geodesic segment which is completely
 contained in ${\cal U}_p$ and connects $q$ with $q'$. \\
Statements and proofs of existence 
 of normal, totally normal  and  convex neighborhoods of 
any point of any geodesically complete manifold 
can be found in  \cite{kn} for affine connections  and \cite{dc,bee} 
for the Riemannian and Lorentzian case respectively.
If a (complete) 
Riemannian manifold has an {\em injectivity radius} $r>0$ \cite{dc}, then
each pair of points $p,q$ with $d(p,q)<r$ 
is contained in a totally normal neighborhood. 

If ${\cal M}$ admits a boundary, all the definitions above
and results concerning normal, totally normal and convex neighborhoods 
of points away from the boundary hold true.\\

\noindent {\bf 1.2}
{\em Heat-kernel coefficients, Hadamard parametrix and the problem
of the symmetry in the arguments.}
In this part we discuss informally some features of heat-kernel coefficients
in both the Lorentzian and the Riemannian case.

In our general hypotheses on the manifold ${\cal M}$ endowed with the metric
${\bf g}$, fixing  any open totally normal (or geodesically 
convex) neighborhood
 ${\cal N}$, the so-called {\em world function} is defined, for $(x,y)\in 
{\cal N}\times{\cal N}$, as the real-valued map
\begin{eqnarray} 
(x,y)\mapsto \sigma(x,y) :=
 \frac{1}{2}{\bf g}(x)(exp_x^{-1}(y), exp_x^{-1}(y)) 
(=  \frac{1}{2}{\bf g}(y)(exp_y^{-1}(x), exp_y^{-1}(x)) )
\:. \label{sigma}
\end{eqnarray} 
$\sigma(x,y)$ does not depend on the chosen  particular
open totally normal neighborhood which contains $x$ and $y$.
As is well-known \cite{kn,bee,dc}, $\sigma$ is always smooth in $(x,y)$ and 
furthermore  analytic in $x$ and $y$ (separately in general) whenever
the metric is analytic. 
 This is because, in open totally normal neighborhoods, the function
$(x,y)\mapsto exp_x^{-1}(y)$ is always $(x,y)$-$C^\infty$
or  analytic in $x$ and $y$ \cite{kn} if the metric is so.
Moreover, 
whenever the metric is Riemannian and the manifold has an
injectivity radius $r>0$ (this holds for  compact manifolds in particular), 
$\sigma$ can be defined on its natural domain 
${\cal D}_r:=\{ p,q \in {\cal M} \:|\: d(p,q)< r\}\label{Dr}$,
 $d$ being the Riemannian distance on the manifold.
Indeed, in the considered situation,
 $\sigma$ belongs to $C^\infty(\{ p,q \in {\cal M} \:|\: d(p,q)< r\})$.
This is because of Sobolev's Lemma \cite{ru}, since, in the Riemannian case,
the function $(x,y)\mapsto \sigma(x,y) = d^2(x,y)/2$ is continuous
everywhere on ${\cal M}\times{\cal M}$ and is smooth in each variable
separately on ${\cal D}_r$.

We pass to summarize
 the main features of the ``small $t$  expansion'' of the heat
kernel $K(t;x,y)$ of the positive  operator $A_0$ in compact Riemannian 
 manifolds supposing that  our
general hypotheses 
 are fulfilled \cite{ch,gi,sh,de,ca,fu,taylor,m1}. 
The heat kernel is  the  integral
 kernel of the semigroup of positive  self-adjoint operators $e^{-tA}$,
$t\in ]0,+\infty[$  which 
is a
solution in $C^{\infty}((0,+\infty)\times{\cal M}\times{\cal M})$
of the ``heat equation'' 
with respect to the operator $A_0$
\begin{eqnarray}
\frac{\partial\:\:}{\partial t} K(t,x,y) + A_{0x} K(s,x,y) = 0\:, \label{b}
\end{eqnarray}
with the initial condition in $C^\infty({\cal M})$
\begin{eqnarray}
\lim_{t\rightarrow 0^+} K(t,x,y) &=& \delta(x,y)\:.  \label{d}
\end{eqnarray}
Fixing  a sufficiently small open geodesically convex  neighborhood
 of the manifold ${\cal N}$\footnote{Actually, with small changes,
a very similar decomposition
 holds in the set ${\cal D}_r$ given above \cite{m1}.},
the ``heat-kernel  expansion at $t\to 0^+$'' is the decomposition of the
heat kernel
\begin{eqnarray}
K(t; x,y) =
\frac{e^{-\sigma(x,y)/2t}}{(4\pi t)^{D/2}}
  \sum_{j=0}^N a_j(x,y) t^j +
\frac{e^{-\eta\sigma(x,y)/2t}}{(4\pi t)^{D/2}} t^N O_{\eta,N}(t;x,y)
\label{expansion1}\:,
\end{eqnarray}
which holds for  $x,y\in {\cal N}$. In (\ref{expansion1}), 
$\eta$ is a real which is arbitrarily fixed in $]0,1[$ and
the function $O_{\eta,N}$ satisfies
\begin{eqnarray}
|O_{\eta,N}(t;x,y)| <C_{\eta,N}|t| \:, \label{lim}
\end{eqnarray}
 uniformly in $(x,y)$. Above, $C_{\eta,N}\geq 0$ does not depend on $t$.
Finally, 
the coefficients $a_j(x,y)$ are smooth functions defined in ${\cal N}
\times{\cal N}$ by  recurrence 
relations we shall examine shortly (see \cite{m1} and the appendix 
of \cite{m2} for details). Similar expansions have been studied extensively
in physics and mathematics and have been generalized  
considering Laplace-like operators acting on smooth sections of vector bundles
on Riemannian/Lorentzian manifolds (see \cite{avramidi} and references 
therein).  For $x\neq y$, the expansion above is not a proper asymptotic 
expansion because the  remaining 
\begin{eqnarray}
R_N(t;x,y) 
:= \frac{e^{-\eta\sigma(x,y)/2t}}{(4\pi t)^{D/2}} t^N O_{\eta,N}(t;x,y)\:,
\label{resto}
\end{eqnarray}
in principle, may be {\em less infinitesimal} than  previous terms
in the expansion in spite of vanishing  faster than any positive power of $t$
as $t\to 0^+$. 
Taking the limit as $\eta \to 1^-$ on the right-hand side of 
(\ref{resto}), one gets
\begin{eqnarray}
R_N(t;x,y)
:= \frac{e^{-\sigma(x,y)/2t}}{(4\pi t)^{D/2}} t^N O_{N}(t;x,y)\:.
\label{resto'}
\end{eqnarray}
However, there is no guarantee that $O_{N}(t;x,y)$ vanishes or is bounded 
as $t\to 0^+$.
Therefore, as said above, 
 the remaining of the formally ``asymptotic'' expansion of $K(t;x,y)$
could be less infinitesimal than the previous terms in the expansion 
(\ref{expansion1}). 
The lack of
general 
information of the precise behaviour of the remaining of the considered 
expansion around $t=0$ does not allow one to get  important
theorems such as the {\em uniqueness} of the coefficients $a_j(x,y)$.
It is worthwhile stressing that, by the symmetry of $K(t;x,y)$ in the
 Riemannian case
and the general symmetry of $\sigma(x,y)$, the symmetry of the coefficients
$a_j(x,y)$ would follow from the  uniqueness theorem trivially.  
Actually,  at least to the author's knowledge, 
 there is no  proof of the general {\em off-diagonal} asymptoticness of the 
heat-kernel expansion in the mathematical literature\footnote{Unfortunately, 
the  important textbook \cite{ch}
reports a  result concerning this point which does not seem to follow
from the corresponding proof (see  Appendix
of \cite{m2}).}. Conversely, there appear 
formulae  concerning upper bounds of the heat
 kernel which  contain some arbitrary parameter like $\eta$ above \cite{de}.
On the other hand, in practice, 
computations concerning {\em off-diagonal} heat-kernel coefficients in 
nontrivial cases are impossibly complicated and therefore, no counterexample
is known concerning their symmetry. 
It is worthwhile remarking that symmetrized expansions for $K(t;x,y)$ 
can be obtained following
different approaches as the ``Weyl calculus''  \cite{taylor}.
However,
the coefficients obtained by this route satisfy {\em different} equations
from the heat-kernel recurrence relations and, in general, cannot
be identified with the standard heat-kernel coefficients used in physics.
Obviously, in the case $x=y$, when both exponentials disappear, the
 heat-kernel expansion (\ref{expansion1}) is a {\em  proper}
 asymptotic expansion.

Whenever ${\cal M}$
is Riemannian and  noncompact, 
the heat kernel  exists  as a smooth function
(see \cite{de,wald79} for the pure Laplacian case) and
the expansion (\ref{expansion1}) 
remains true, in general, provided the injectivity radius
of the manifold is strictly positive (this can be assured by imposing 
bounds on the sectional curvature of the manifold)
and supposing that 
some bound conditions on the Ricci curvature tensor are satisfied
\cite{ch}.

In the presence of boundaries of the Riemannian manifold ${\cal M}$,
  $A$ being some self-adjoint extension of $A_0$
determined by fixing some boundary conditions on $\partial {\cal M}$,
the expansion above has to be changed
  just by adding  a  further (dependent on the boundary conditions) 
term $h(t;x,y)$ in the sum above. However, for $x\neq y$,
 the literature on this case is 
not very extensive, except for the analysis of the pure Laplacian case 
 with Dirichelet boundary conditions. 
In this case  \cite{ch} $h(t;x,y)$ 
can be bounded by a constant times
$t^{D/2} e^{-\sigma(y,\partial{\cal M})/4t}$ (or $x$ in place of $y$) 
 and thus vanishes exponentially
as $t\to 0^+$ whenever at least one of the arguments does not belong to 
the boundary. 	In the case $x=y$, also $h(t;x,x)$ can be expanded 
in an asymptotic series   of terms.  
These terms  carry powers of the form
$t^{j+1/2-D/2}$ instead of $t^{j-D/2}$ ($j$ natural) \cite{el,ze}
 and maintain the exponential factor cited above.
Hence, in the case $x=y$ away from the boundary,
these added terms 
vanish faster than any power $t^M$ ($M\in \N$) as $t\rightarrow 0^+$
(see \cite{ch} for the pure Laplacian case).

In the Lorentzian case, the picture changes dramatically. Generally, $A_0$
is not bounded below and this drawback in general
remains in  self-adjoint extensions\footnote{They exist because 
$A_0$, thought as an operator on $L^2({\cal M}, d\mu_g)$,
is symmetric ( e.g., on $C^\infty_0({\cal M})$) and  
it commutes with the antiunitary operator given by the complex conjugation
\cite{rs}.}.
This introduces several pathologies dealing
with the heat equation and the associated semigroup of exponentials
which, as a  consequence, contains unbounded operators. 
However, an analogous
expansion should arise considering an  ``heat kernel'' $H(s,x,y)$
solution of a ``Schr\"{o}dinger'' equation \cite{bd,fu} formally related
to the group of the {\em imaginary} exponentials operators 
(which are bounded) 
of the operator $A_0$, 
\begin{eqnarray}
-i \frac{\partial\:\:}{\partial s} H(s,x,y) + A_{0x} H(s,x,y) = 0\:; \label{a}
\end{eqnarray}
with initial condition (holding on locally-integrable 
smooth test functions)
\begin{eqnarray}
\lim_{s\rightarrow 0} H(s,x,y) &=& \delta(x,y)\:.  \label{c}
\end{eqnarray}
(See \cite{fu,ca,avramidi} 
for details.)
In the Lorentzian case, one expects that  some local
``asymptotic'' expansion of the form\footnote{We stress that, if
 manifold is Lorentzian, $\sigma(x,y)$ can also be negative.}
\begin{eqnarray}
H(t,x,y) \sim
\frac{e^{i\sigma(x,y)/2s}}{(4\pi is)^{D/2}}
  \sum_{j=0}^\infty a_j(x,y) (is)^j
\label{expansion2}\:
\end{eqnarray}
should hold. If the manifold has a boundary, further terms appear and depend
on the boundary.
Actually, the situation is much more complicate \cite{fu,esfu} and we shall
not address this issue here. We only notice that, if the
Lorentzian manifold is locally {\em static} and 
$V$ is invariant under the associated  group of isometries,
 it should be possible 
to get information on the Lorentzian heat-kernel coefficients by
a Wick-rotation into a Riemannian manifold. In this case, 
the analytical dependence  of the heat-kernel coefficients on  the
time associated to the time-like Killing vector should be {\em a consequence}
of the staticity of the metric. This should allow to perform the
analytical continuation to
 the Euclidean time.\\

\noindent {\bf 1.3} {\em Determination and smoothness 
of heat-kernel and Hadamard coefficients}.
Let us consider a manifold ${\cal M}$ satisfying our general hypotheses.
In a local coordinate system $x^1,x^2,\cdots, x^D$,
defined in  any open convex neighborhood or, more generally, 
in any open totally normal neighborhood ${\cal T}$,
one can define the  van Vleck-Morette determinant
$\Delta_{VVM}$. This is a {\em bi-scalar} which is 
given in the coordinates above by
\cite{ca}
\begin{eqnarray}
\Delta_{VVM}(x,y) &:=& -\left[g(x)g(y)\right]^{-1/2}
\det \left( \frac{\partial^2 \sigma(x,y)}{\partial x^{a} \partial y^{b}}
 \right) > 0
\label{vvmdef}\:,
\end{eqnarray}
$x,y \in {\cal T}$.
On ${\cal T}$, 
it satisfies (all derivatives are computed in the variable $x$ in the 
considered coordinate system)
\begin{eqnarray}
\nabla^a\sigma(x,y)\nabla_a\ln \Delta_{VVM}(x,y) = -D + \nabla_c\nabla^c 
\sigma(x,y) \:. \label{stra}
\end{eqnarray}
Notice that $\Delta_{VVM}(x,y)$ is strictly
 positive on ${\cal T}$ (since it is a bi-scalar and 
$\Delta_{VVM}(x,y) = |g(x)|^{-1/2}>0$ 
in normal coordinates around $y$) and it is
a $C^{\infty}$ function of $(x,y)$
which is also analytic in $x$ and $y$ (separately in general)
whenever the manifold and the metric are $C^\infty$ and analytic respectively.
Obviously, this follows from the fact that $(x,y) \mapsto \sigma(x,y)$
is $(x,y)$-$C^\infty$ or, respectively, analytic in $x$ and $y$ in the 
considered domain. If the manifold is Riemannian and 
 the injectivity radius is strictly positive, 
the van Vleck-Morette determinant
can be defined on the set ${\cal D}_r$ as a smooth function.

In either Riemannian or Lorentzian  case, 
the coefficients $a_j$ are bi-scalars defined
 in any fixed open geodesically  convex  neighborhood ${\cal N}$
containing $x$ and $y$,
away from the boundary of the manifold if it exists,
or equivalently, in the set ${\cal D}_r$, provided the manifold is 
Riemannian with  strictly positive injectivity
 radius.
In the considered domain, the functions 
$(x,y) \mapsto a_j(x,y)$ can be heuristically determined by well-known
 equations with opportune regularity conditions.
These equations
 are obtained perturbatively inserting the considered 
Lorentzian or Riemannian expansions formally computed up to 
$N=\infty $ (omitting the remaining) 
into (\ref{a}) and (\ref{b})  and imposing that each coefficient of 
any power of $t$ vanishes separately and taking into account (\ref{stra}). 
Following this route, in any normal coordinate system defined 
into a normal neighborhood of  $y$, one finds
\begin{eqnarray}
 \frac{d}{d\lambda} \left(a_0(x(\lambda),y) 
\Delta_{VVM}^{-1/2}(x(\lambda),y)\right) &=& 0
\label{a0}\:,\\
- \lambda^j 
\Delta_{VVM}^{-1/2}(x(\lambda),y)\: A_{0x(\lambda)} 
a_{j}(x(\lambda),y)  &=&
  \frac{d}{d\lambda}\left( \lambda^{j+1}
a_{j+1}(x(\lambda),y)\: \Delta_{VVM}^{-1/2}(x(\lambda),y)\right)
  \label{coefficienti}\:,
 \end{eqnarray}
where $\lambda \mapsto  x(\lambda)$ is the unique segment
geodesic from $y \equiv x(0)$ 
to $x \equiv x(1)$ completely 
contained in the normal neighborhood.\\
The regularity conditions are the following ones.
The solutions have to be 
 $(x,y)$-smooth  everywhere in the considered domain, 
in particular they have to be   bounded for $x\rightarrow y$.
Moreover, 
\begin{eqnarray}
a_0(x,y) \rightarrow 1
\end{eqnarray}
must be hold for $x \rightarrow y$, which  assures the 
validity of (\ref{c}) and (\ref{d})
 since also $\Delta_{VVM}^{-1/2}(x,y) \rightarrow 1$. \\
The reason for using open geodesically convex
neighborhoods should be clear.
Indeed, in order to perform the derivatives contained in the differential 
 operator  $A_0$ on the left-hand side 
of (\ref{coefficienti}), with $x$
and $y$ fixed in ${\cal N}$, 
there must exist an open neighborhood ${\cal O}_z$ of  any point
$z$ which belongs to the geodesic which connects $y$ with $x$, such that
a geodesic which connect $y$ with any point in ${\cal O}_z$ lies
completely in ${\cal N}$. Moreover the dependence of the considered geodesics 
on  the extreme points has to be smooth. This is true  provided 
 ${\cal N}$ is open and geodesically convex,
the smoothness
being a consequence of the  total normality of the neighborhood. 
(Working in ${\cal D}_r$ whenever possible, similar properties hold true
and the definitions are well-posed.)\\
The following definition gives
the unique solutions in ${\cal N}$  
 of the recurrence equations (\ref{a0}) and 
(\ref{coefficienti}) ($j\geq 1$) satisfying  the requirements given above, 
either for Riemannian and Lorentzian manifolds.\\

\noindent {\bf Definition 2.1.} {\em In our general hypotheses on 
${\cal M}$ and $A_0$, where the former can admit a boundary and can be
either Riemannian or Lorentzian, in any  fixed open geodesically convex 
neighborhood ${\cal N}$ not intersecting $\partial {\cal M}$,
the heat-kernel coefficients are 
the real-valued functions defined on  ${\cal N} \times {\cal N}$, 
labeled
by $j\in \N$,
\begin{eqnarray}
a_{0}(x,y) &=&  \Delta_{VVM}^{1/2}(x,y) 
\label{vvm}\:,\\
a_{(j+1)}(x,y) &=& - \Delta^{1/2}_{VVM} (x,y) \int_0^{1}
\lambda^{j}\left[  \Delta^{-1/2}_{VVM} A_{x(\lambda)} a_j\right] 
(x(\lambda), y) d\lambda\:,
\label{s}
\end{eqnarray}
$\lambda \mapsto  x(\lambda)$ being the unique 
geodesic segment from $y \equiv x(0)$ 
to $x \equiv x(1)$ contained completely in ${\cal N}$.}\\ 

(It is possible to give an analogous and equivalent
 definition on the set
${\cal D}_r$ in a Riemannian manifold with strictly positive injectivity
radius. In any case, it is obvious that,
 fixing $x,y$, $a_j(x,y)$ defined above 
does not depend on the chosen 
 open geodesically convex neighborhood containing $x$ and $y$).

In the case of a Riemannian compact manifold, 
the heat kernel coefficients defined above  are just those which appear in 
(\ref{expansion1}) \cite{ch,m1,m2}. Moreover, these coefficients 
do not depend on the particular self-adjoint extension of $A_0$.
(Conversely, in the case of the presence of a boundary, the further 
coefficients cited previously {\em do depend} on the self-adjoint extension).\\
$a_0(x,y)$ enjoys the same properties of positivity, smoothness/analyticity 
of $\Delta_{VVM}(x,y)$.
Moreover, assuming the smoothness/analyticity of the function $V$ 
which appears  in the operator $A_0$ and  
working in  local coordinates defined in a    geodesically convex
 neighborhood containing $x$ and $y$, one can generalize this result
to all the coefficients $a_j$. Indeed,
 taking account of the 
smooth/analytic  dependence on  the parameter and the initial and final 
conditions of the geodesics
(and their derivatives) \cite{dc,kn} and finally
considering (\ref{s}),
 one can check that the coefficients $a_j(x,y)$ are $(x,y)$-smooth or
$x$ and $y$ analytic functions
of $(x,y)$ in the considered domain (away from the 
boundary). Concerning the proof of analyticity, a short 
way is to continue the functions on the right-hand side of (\ref{s})
to complex values of the arguments $x$ and $y$. For example, since
the integrand  functions are analytic in $x$ for $y$ fixed, one
 can continue these
to complex
values in the variable $x$ for any  fixed real $y$. Therefore one can prove
Cauchy-Riemann's conditions for the complex components of 
$x$ on  the left-hand side of (\ref{s})
 passing the derivatives under the sign of integration.

Also in the presence of boundaries,
 one can formally use the coefficients $a_j$ above (those which do not depend
on the boundary conditions)
to build up a part of a formal series 
 for  Green's functions  $G(x,y)$
\begin{eqnarray}
A_{0x} G(x,y) = \delta(x,y)\:.
\end{eqnarray}
Indeed one has  that the Green functions above can be locally approximated 
by formal series, which defines, whenever they converge to proper solutions,
``Hadamard local fundamental solutions''. Both in the Riemannian  
and  in the  Lorentzian case,
these series  
 can be represented, {\em up to the indicated order}, by 
(the summation appears for $D \geq 4$ only)
\begin{eqnarray}
H(x,y)
&=& \sum_{j=0}^{D/2-2} (D/2-j-2)!\left(\frac{2}{\sigma}\right)^{D/2-j-1}
\frac{a_{j}(x,y|A)}{(4\pi)^{D/2}}\nonumber\\
& & -\frac{2a_{D/2-1}(x,y|A)
- a_{D/2}(x,y|A) \sigma}{2(4\pi)^{D/2}} \ln \left(\frac{\sigma}{2}\right)
 \label{unop}
\end{eqnarray}
if $D$ is even, and (the summation appears for $D\geq 5$ only)
\begin{eqnarray}
H(x,y) &=& \sum_{j=0}^{(D-5)/2}
\frac{(D-2j-4)!! \sqrt{\pi}}{2^{(D-3)/2-j}}
\left(\frac{2}{\sigma}\right)^{D/2-j-1}
\frac{a_{j}(x,y|A)}{(4\pi)^{D/2}} \nonumber \\
 & & +   \frac{a_{(D-3)/2}(x,y|A)}{(4\pi)^{D/2}}
\sqrt{\frac{2\pi}{\sigma}}
-  \frac{a_{(D-1)/2}(x,y|A)}{(4\pi)^{D/2}}
\sqrt{2\pi\sigma}
\label{unop'}
\end{eqnarray}
if $D$ is odd, see \cite{m2}.\\
In fact, (\ref{coefficienti}) and (\ref{vvm}) assure
that the coefficients of the
formal series for the considered Green's function, truncated
at the indicated orders, satisfy the corresponding recurrence differential
equations given in Chapter 5 of
\cite{garabedian} both in the Riemannian and in the Lorentzian case
and the  corresponding regularity/initial conditions.
For $D$ even, the non-divergent part of the Hadamard expansion, omitted above, 
is ambiguous and depends on the choice of a coefficient. However, in the
point-splitting technique, the symmetry is requested for the coefficients 
which appear above only. 
In the Lorentzian case one has to specify the prescription to compute 
the logarithms and the fractional powers of $\sigma$ in the case $\sigma<0$;
as is well known in Quantum Field Theory,
 this produces differnt types of (Hadamard expansions of)
two-point functions with the same coefficients (Wightman functions and
Feynman propagator).
The expansions above
 define  ``parametrices'' of  the Green's functions  at the
considered order of approximation.
However, nothing  assures that the  corresponding {\em not truncated} series 
converge and, most important, define
smooth functions. This convergence does
anyway hold locally uniformly  in the Riemannian case
\cite{garabedian}, in this case the series define proper functions. 
Furthermore,
  if one deals with a real section of a complex
analytic manifold where the complex analytic function $V$ is also real, then
 the sum of the series is also analytic and thus 
smooth (by standard theorems of complex
function theory) and define a true local solution of the
corresponding differential equation: these are the Hadamard fundamental 
local solutions. The Lorentzian case, 
in particular cases at least, can be treated 
by analytic-continuation procedures
from the Riemannian case, but, in general, one has a convergence
in the sense of Borel only \cite{fu,fr}.  However,
all  these issues should not be
so important in  practice,
since, within the practical point-splitting procedure, one has to take into
account only  a finite number of terms of these expansions and thus one
can  use the parametrices instead of the sum of the series. Nevertheless
the requirements above  on the convergence of the series
as well as the smoothness of the sum
have been used within the proof of the conservation of the stress
tensor \cite{wald78,b,bo}. These requirements have been partially dropped
in \cite{fsw} where a ``distributional'' convergence of the Hadamard
series have been used, but no explicit improved proof of the results 
given in \cite{wald78} (and related papers) have been supplied. 

As we said previously, 
another strongly 
important point, used to prove the conservation of the renormalized 
stress tensor in the cited literature, is the {\em symmetry}
 of the Hadamard local fundamental solutions concerning  their
divergent part as $\sigma \to 0$ \cite{wald78} up to the order of expansion 
considered in (\ref{unop}) and (\ref{unop'}) (actually the most part
of the known literature treats the case $D=4$ only, but the same 
procedures can by generalized to different dimensions in a direct way).
In \cite{fsw}, it was argued that a proof of this property 
holds true also for the case of a $C^{\infty}$ (not analytic) manifold, 
unfortunately such a proof 
was not reported there  and, at least to the author's knowledge, 
such a general proof (as well as a proof of the symmetry of the 
heat-kernel coefficients) does not exist in the literature.
Notice that the symmetry of the heat-kernel coefficients assures the
symmetry of the parametrices (\ref{unop}) and (\ref{unop'}). For this reason 
the symmetry of the heat-kernel coefficients is important in the
point-splitting technique.

\section{A proof of the symmetry of heat kernel coefficients in the Riemannian
case.}

{\bf 2.1} {\em Two theorems.} Our proof is quite technical and involves 
several steps. The way  is dealt with as follows. First,
one shows that the thesis holds true in the case of a real analytic manifold
by using known local properties of the expansion of the heat-kernel.
This is the content of the first theorem we shall prove.
Afterwards, one proves that, in some sense, any $C^\infty$ manifold can be
approximated by analytic manifolds. This point is quite complicated
because this approximation has to hold into a common geodesically 
convex  neighborhood. This is necessary 
in order to make sensible a common definition of heat-kernel
coefficients. Finally, one proves that the heat-kernel coefficients, defined
in the common geodesically  convex 
neighborhood are ``sequentially continuous'' 
in the class of the used metrics. Then, and this is the content of the second
theorem we shall present, the symmetry for the case of a $C^\infty$ manifold
follows by the ``continuity'' of the heat-kernel coefficients with respect
to the metrics and from the symmetry in the analytic case.
It is worthwhile stressing that (local and global) 
approximation theorems in real analytic
 manifolds are well-known in the literature (see \cite{to} for a recent 
review). However, these theorems concern functions rather than metrics and
the problem of the existence of common geodesically convex neighborhoods
is not treated explicitly. For this reason we prefer giving independent 
proofs (see the appendix of this work).
\\

\noindent{\bf Lemma 2.1.}
{\em Let us assume our general hypotheses on $A_0$ and 
 ${\cal M}$ which is explicitly supposed to be Riemannian
and compact.}
{\em In a coordinate system defined in an open sufficiently
small (geodesically convex) neighborhood ${\cal N}_z$ of any point 
$z\in {\cal M}$, 
for any pair of points $x,y\in {\cal N}_z$, and
any natural $N$ such that
$N> D/2 +2|\alpha'| + 2|\beta'|$, $\alpha'$, $\beta'$ being arbitrarily 
fixed multi-indices, one has
\begin{eqnarray}
D^{\alpha'}_x D^{\beta'}_y 
K(t;x,y) &=& D^{\alpha'}_x D^{\beta'}_y 
\left\{  \frac{e^{-\sigma(x,y)/2t}}{(4\pi t)^{D/2}}
\sum_{j=0}^{N} a_j(x,y) t^j\right\}\nonumber\\
&+&  \frac{e^{-\eta\sigma(x,y)/2t}}{(4\pi t)^{D/2}}
t^{N-|\alpha'|-|\beta'|} O_{\eta,N}^{(\alpha',\beta')}(t;x,y)\:. \label{lunga}
\end{eqnarray}
Above, $\eta\in ]0,1[$ can be fixed arbitrarily and the corresponding
 function 
$O_{\eta,N}^{(\alpha',\beta')}$ is  continuous in $(t,x,y)\in [0,+\infty[
\times{\cal N}_z\times{\cal N}_z$ and
$(x,y)$-uniformly bounded  by $B_{\eta,N}|t|$
in a positive neighborhood of $t=0$, $B_{\eta,N}>0$ being a constant.}\\

\noindent {\em Proof.}  See {\bf Lemma 2.1} of \cite{m2}.\\

\noindent By the lemma above we are able to prove our first result.\\

\noindent {\bf Theorem 2.1.}
{\em In our general hypotheses on ${\cal M}$, which is supposed to be 
a Riemannian manifold (also with boundary in general), 
and $A_0$,  the following properties hold true for the
heat-kernel coefficients in} (\ref{vvm}) {\em and} (\ref{s}). 

(a) {\em Fixed any point $z\in {\cal M}$ (away from $\partial{\cal M}$),
there is a sufficiently small open geodesically convex 
neighborhood of $z$, ${\cal N}_z$ 
(which does not intersect $\partial {\cal M}$),
such that,  in any local coordinate 
system defined therein, for any $j\in \N$,
and  any pair of derivative operators $D_x^\alpha$, $D_x^\beta$ and for any
 point $y\in {\cal N}_z$}
\begin{eqnarray}
D_x^\alpha D_y^\beta
 a_j(x,y)|_{x=y} = D_x^\alpha D_y^\beta a_j(y,x)|_{x=y}
\label{prima}\:.
\end{eqnarray}

(b) {\em For any choice of the multi-indices $\alpha,\beta$ and $j,N\in \N$,
the functions
\begin{eqnarray}
\Omega_j(x,y) := a_j(x,y) - a_j(y,x) \label{omegaj}  
\end{eqnarray}
 computed in any local coordinate system in the set ${\cal N}_z$
defined in} (a),
{\em satisfy}
\begin{eqnarray}
\sigma^{-N}(x,y)  \left[D_x^\alpha D_y^\beta \Omega_j(x,y)\right]
\rightarrow 0
\end{eqnarray}
{\em as $x\rightarrow y$.} 

(c) {\em If $g_{ab}$ and $V$ are (real) analytic functions of some
local coordinate frame defined in an open connected set
${\cal O}$ (away from the boundary) where, for all $x,y\in {\cal O}$,
the coefficient $a_j$ ($j$ fixed in $\N$) is defined, then
\begin{eqnarray}
a_j(x,y) = a_j(y,x) \label{strong}
\end{eqnarray}
for any pair $(x,y) \in {\cal O}\times {\cal O}$.}\\

\noindent {\em Proof.} 
Notice that the thesis  
 is trivially proven for $a_0(x,y) = \Delta_{VVM}^{1/2}(x,y)$,
since the right-hand side is symmetric in $x$ and $y$. So we can pass
directly to the case $j>0$  proving (a).
Let us first consider the case of a compact 
 Riemannian manifold. Then, since $A_0$ is positive, we 
can employ standard theorems on the heat kernel, in particular 
 we can use  {\bf Lemma 2.1} above. Therefore, let us
fix a  coordinate system where  {\bf Lemma 2.1} holds true 
in an open geodesically convex neighborhood of the point $z$. 
For any pair of multi-indices $\alpha, \beta$,
\begin{eqnarray}
D_x^\alpha D_y^\beta 
\sum_{j=0}^{N} \Omega_j(x,y) t^j
= (4\pi t)^{D/2} D_x^\alpha D_y^\beta \left\{ e^{+\sigma(x,y)/2t} 
\left[ \frac{e^{-\sigma(x,y)/2t}}{(4\pi t)^{D/2}} 
\sum_{j=0}^{N} \Omega_j(x,y) t^j\right] \right\}\nonumber\:.
\end{eqnarray}
Taking into account that $(x,y)\mapsto K(t;x,y) - K(t;y,x) \equiv 0$ 
in these hypotheses
\cite{ch} (see also {\bf Theorem 1.1} in \cite{m1}),  employing 
Leibnitz' rule in evaluating the derivatives above and making use of
(\ref{lunga}), we get
\begin{eqnarray}
D_x^\alpha D_y^\beta 
\sum_{j=0}^{N} \Omega_j(x,y) t^j=
 t^{N-2|\alpha|-2|\beta|} e^{\delta \sigma(x,y)/2t} 
U^{(\alpha,\beta)}_{\delta,N}(t;x,y) \:,
\end{eqnarray}
where $\delta = 1-\eta \in ]0,1[$ ($\eta$ is the same parameter
which appears in (\ref{lunga})), $U^{(\alpha,\beta)}_{\delta,N}(t;x,y)$
is built up using  linear combinations of antisymmetrized remainings which appear
in (\ref{lunga}), 
$O_{\eta,N}^{(\alpha',\beta')}(t;x,y) -O_{\eta,N}^{(\alpha',\beta')}(t;y,x) $
with coefficients given by positive powers of $t$ and derivatives of the
function $\sigma$. 
Due to 
the similar property of the functions 
$O_{\eta,N}^{(\alpha',\beta')}$,
  $(t;x,y) \mapsto U^{(\alpha,\beta)}_{\delta,N}$
is $(x,y)$-uniformly
 bounded by some constant $C_{\delta,N}>0$ in a right-neighborhood
of $t=0$,  provided $N$ has been  chosen 
large sufficiently. 
Then, taking the limit for $x\rightarrow y$ we have
\begin{eqnarray}
\sum_{j=0}^{N}  D_x^\alpha D_y^\beta \Omega_j(x,y)|_{x=y} t^j
= t^{N-2|\alpha|-2|\beta|} U^{(\alpha,\beta)}_{\delta,N}(t;y,y) \:,
\end{eqnarray}
and thus, with a trivial redefinition of $U$ obtained by decomposing
\begin{eqnarray}
\sum_{j=0}^{N} =  
\sum_{j=0}^{N -2|\alpha| -2|\beta| -1} +\sum_{N -2|\alpha| -2|\beta|}^{N}\:,
\nonumber 
\end{eqnarray}
one gets
\begin{eqnarray}
\sum_{j=0}^{N -2|\alpha| -2|\beta| -1}  D_x^\alpha D_y^\beta 
\Omega_j(x,y)|_{x=y} t^j
= t^{N-2|\alpha|-2|\beta|}
 V^{(\alpha,\beta)}_{\delta,N}(t;y,y) \:,
\end{eqnarray}
where  $V^{(\alpha,\beta)}_{\delta,N}(t;y,y)$ is bounded in a positive 
neighborhood of $t=0$.
In the limit $t\rightarrow 0^+$,
this is possible only when all the coefficients of the polynomial on the 
left-hand side vanish separately. This implies that
 also the covariant derivatives
of any order of the functions $\Omega_j$, evaluated on the diagonal,
vanish. Obviously this does not depend on the particular coordinate frame
used around $y$. Therefore, changing coordinates and passing
from covariant derivatives to ordinary derivatives in a different coordinate 
frame, we get that, once again, the derivatives of any order
of the functions $\Omega_j$, evaluated on the diagonal vanish.
(a) has been proven in the hypotheses of 
a compact 
manifold (without boundary). Given a general manifold 
${\cal M}$ and any inner point $y$, we can consider a neighborhood ${\cal O}$
of $y$ and build up a new manifold ${\cal M}'$ which contains a neighborhood
${\cal O}'$ 
isometric to ${\cal O}$.
 ${\cal M}'$ can be chosen compact (without boundary)
provided ${\cal M}$ is complete. On ${\cal M}'$, we can define an operator
$A'_{0}$ (depending on a smooth  potential $V'$)
which coincides with $A_0$ in the neighborhood ${\cal O}'\equiv {\cal O}$.
In general, also if $A_0$ is positive,
 $A'_0$ may be non-positive.
However, since $V'$ is bounded below by some real $v$, the operator
$A'_0 + |v| I$ is positive on ${\cal M}'$. We can consider the heat-kernel
coefficients $b_j(x,y)$
 of the expansion (\ref{expansion1}) for the operator $A + |v| I$. For these 
coefficients the item (a) of the thesis holds true. An algebraic computation
based on the fact that, formally, if $S(t;x,y)$ satisfies the heat equation
with respect to $A_0$ then  $S(t;x,y) \exp{(-ct)}$ satisfies the same 
equation with respect to $A_0 +cI$ ($c\in \R$),
proves that the coefficients $a_{j}(x,y)$ of (\ref{vvm}) and (\ref{s})
corresponding to $A'_0\equiv A_0$ are related to those above by the relations
\begin{eqnarray}
a_j(x,y) &=& \sum_{k=0}^{j}\frac{(-1)^k |v|^k b_{j-k}(x,y)}{k!}\label{bb}\:,\\
b_j(x,y) &=& \sum_{k=0}^{j}\frac{ |v|^k a_{j-k}(x,y)}{k!}\label{b'}\:.
\end{eqnarray}
The coefficients on the right-hand side of (\ref{bb})
satisfy (\ref{vvm}) and 
(\ref{s}) with respect to $A_0 \equiv A'_0$ in ${\cal O}\equiv {\cal O}'$
once the coefficients on the right-hand side 
do so with respect to $A'_0-|v|I$. Thus, the item (a) is  trivially 
proven for the coefficients $a_j(x,y)$ in the general case. Notice that,
in the same way,
also items (b) and (c) hold true in the general case provided
they are valid  in the particular case of a compact 
 manifold without boundary.\\
 The item (b) is trivially proven by expanding, in the variable $x$,
 any $\Omega_j(x,y)$ and all of its derivatives,
 via Taylor algorithm, around the point $y$ in a normal
Riemannian coordinate system centered in $y\equiv 0$. 
For instance, considering $\Omega_j$, one has, for any $N\in \N$
\begin{eqnarray}
\Omega_j(x,y) = \sum_{0\leq |\alpha|\leq 2N+1 }
\frac{(x^{1})^{\alpha_1}\cdots (x^{D})^{\alpha_D}}{\alpha_1! \cdots \alpha_D!}
\frac{\partial^{|\alpha|} \Omega(x,0)}{\partial^{\alpha_1}x^1 \cdots 
\partial^{\alpha_D} x^D}|_{x=0} + |x|^{2N+1} O_{2N+1}(x)\:, \label{taylor}
\end{eqnarray}
where $|x|^2/2 = \sigma(x,y)$, and $O_{2N+1}(x)$ is a smooth function which
vanishes as $x\rightarrow 0\equiv y$ and thus is bounded around $y\equiv 0$. 
Using the result of the item (a) (changing coordinates in general), one
gets the thesis (b). The same procedure can be employed for derivatives
of $\Omega_j(x,y)$. \\
Let us consider the item (c).  In this case, the Taylor expansion above
can be carried out, in the considered coordinates,
 up to $N=\infty$. Thus, taking into account (a), for $x$ belonging 
to a neighborhood of any fixed point $y\in {\cal O}$,  one has  
\begin{eqnarray}
\Omega_j(x,y) = 0\:.
\end{eqnarray}
Since, for $y$ fixed in ${\cal O}$, 
$\Omega_j(x,y)$ is analytic in  $x \in {\cal O}$,
which is an open and connected set, and vanishes in an open 
neighborhood contained in ${\cal O}$ (dependent on $y$),
 it  has to vanish everywhere on ${\cal O}$.
Therefore $(x,y)\mapsto \Omega_j(x,y)$ vanishes in ${\cal O}\times {\cal O}$.
$\Box$\\

The results obtained above concerning the heat-kernel coefficients,
can be generalized directly to the relevant
coefficients $u_j,v_j$ of the expansions of the Hadamard local solution of the
operator $A_0$ by taking into account (\ref{unop}) and (\ref{unop'}) above. 
Actually, the result contained in the item (b)  should be
sufficient for all applications of  (Euclidean) point-splitting procedures
known in the literature despite the complete 
symmetry of the Hadamard coefficients was
originally requested\footnote{I am grateful to R. M. Wald for this remark.}.
However, we aim to get a more general result.\\

\noindent {\bf Proposition 2.1}. {\em Let 
${\cal M}$ be a real $C^\infty$ manifold
 with a non-singular  metric ${\bf g}$, 
satisfying our general hypotheses.}
 
(a) {\em Let $\Omega'$ be any open set in ${\cal M}$
(such that $\Omega' \cap \partial {\cal M} =\emptyset$) endowed 
with a coordinate frame $x^1,\cdots,x^D$. For 
any  connected relatively-compact open set
 $\Omega$, such that 
 $\bar{\Omega} \subset \Omega'$, there is  a sequence of real  
 metrics $\{{\bf g}_n\}$ with the same signature of ${\bf g}$
defined in 
a neighborhood of $\bar{\Omega}$ such that each $g_{nab}$ is an  analytic 
function of the given coordinates  and the 
sequence $\{ {\bf g}_n\}$ 
converges 
uniformly in $\bar{\Omega}$
 to the metric ${\bf g}$. Similarly, for any fixed  multi-index  $\alpha$, 
the sequance of  derivatives with respect to the coordinates $x^1,\cdots,x^D$
$\{ D^{\alpha}{\bf g}_n\}$, 
converges uniformly in $\bar{\Omega}$
to $\{ D^{\alpha}{\bf g}\}$.} 

(b) {\em For any choice of the set $\Omega'$, the coordinates 
$x^1,\cdots,x^D$, the set $\Omega$ and the sequence 
$\{{\bf g}_n\}$ given above and for any $z\in \Omega$,
there is a natural $N_0$ and a family of  open neighborhoods of
 $z$, $\{{\cal N}_{zi}\}$, $i\in \R$, 
such that  $\{{\cal N}_{zi}\}$ is a local base of the topology of ${\cal M}$,
 ${\cal N}_{zi} \subset \bar{\cal N}_{zi'} \subset \Omega$,
 for any pair $i,i'$ such that  $i'>i$ and, moreover,
for any $i\in \R$, both 
${\cal N}_{zi}$ and $\bar{\cal N}_{zi}$ are
common geodesically convex neighborhoods of  $z$  for all 
the metrics ${\bf g}$ and ${\bf g}_n$ when $n>N_0$.}

(c) {\em For any choice of the set $\Omega'$, the coordinates 
$x^1,\cdots,x^D$, the set $\Omega$, the sequence 
$\{{\bf g}_n\}$,  $z\in \Omega$ and the class $\{{\cal N}_{zi}\}$, 
 $i\in \R$ arbitrary, 
the functions
$ (x,y) \mapsto \sigma_n(x,y)$ are well-defined 
and smooth  in any neighborhood of
 $\bar{{\cal N}}_{zi}\times
\bar{{\cal N}}_{zi}$ and the sequence of these functions as well as 
the sequences of their derivatives
 of any order converge uniformly in $\bar{{\cal N}}_{zi}\times
\bar{{\cal N}}_{zi}$  to $\sigma(x,y)$ and corresponding derivatives.}

(d) {\em For any choice of the set $\Omega'$, the coordinates 
$x^1,\cdots,x^D$, the set $\Omega$, the sequence 
$\{{\bf g}_n\}$,  $z\in \Omega$ and the class $\{{\cal N}_{zi}\}$, for any
 $i\in \R$, 
if $(\lambda,x,y) \mapsto \gamma_n(\lambda,x,y)$, $\lambda \in [0,1]$, 
indicates the only  geodesic segments 
starting from the  point $y\in \bar{{\cal N}}_{zi}$ and terminating in
 the  point $x\in \bar{{\cal N}}_{zi}$ corresponding
to the $n$-th metric and contained in 
$\bar{{\cal N}}_{zi}$,
then $\{\gamma_n(\lambda,x,y)\}$ and the sequences of their 
$\lambda,x,y$-derivatives
of any order converge uniformly
in $[0,1]\times \bar{{\cal N}}_{zi}
\times \bar{{\cal N}}_{zi}$  to  
$\gamma(\lambda,x,y)$ 
and corresponding derivatives,
$\gamma(\lambda,x,y)$ being the geodesic of the initial metric ${\bf g}$.}\\

\noindent {\em Proof.} See {\bf Appendix} $\Box$.\\

\noindent We need another technical lemma to get the final theorem.\\

\noindent {\bf Lemma 2.2.}
{\em Let $\{g_{k,n}\}$ be a class of continuous functions,
$k=1,2,\cdots,l$ and $n\in \N \cup \{\infty\}$,} 
\begin{eqnarray}
g_{k,n} : K_k \to M_k \:,
\end{eqnarray}
{\em where $M_k$ and  $K_k \subset M_k$ are, respectively, metric spaces and
 compact sets.
 Let $\{f_{n}\}$ be a class of
continuous functions,  $n\in \N \cup \{\infty\}$,  
\begin{eqnarray}
f_{n} : \Omega_1 \times \Omega_2\times \cdots \times \Omega_l \to N\:,
\end{eqnarray}
where $N$ is a metric space, the  sets $\Omega_k
\subset M_k$, $k=1,2, \cdots, l$,
 are open and $g_{k,\infty}(K_k)\subset \Omega_k$.
Suppose that, for any fixed $k$ and for $n\to +\infty$, 
$g_{k,n} \to g_{k,\infty}$ uniformly in $K_k$ 
and $f_{n} \to f_{\infty}$ uniformly in $\Omega_1\times\cdots \times 
\Omega_l$.
Then, there is a natural $N_0$ such that, for $n>N_0$, the left-hand 
side below is well-defined and, for $n\rightarrow +\infty$, 
\begin{eqnarray}
f_n(g_{1,n}(x_1),g_{2,n}(x_2),\cdots, g_{l,n}(x_l))
\to
f_\infty(g_{1,\infty}(x_1),g_{2,\infty}(x_2),\cdots, g_{l,\infty}(x_l))
\end{eqnarray}
uniformly in $K_1 \times K_2 \times \cdots \times K_l$.}\\

{\em Proof.} It is quite straightforward. 
Take into account that a 
 continuous function $h$ defined on a compact set $H$ of a metric space
with values in a metric space  
is uniformly continuous in $H$ and $h(H)$ is also a compact set. 
$N_0$ is defined by determining
the compact sets 
$C_1,\cdots, C_l$ such that $g_k(K_{k,n}) \subset C_k \subset
\Omega_k$, for $n= N_0+1,N_0+2,\cdots, \infty$.
$\Box$ 
\\

We are now able to state and prove the most important theorem concerning the
symmetry of heat-kernel coefficients in Riemannian manifold.\\

\noindent {\bf Theorem 2.2}. {\em 
Let {\cal M} be a $C^\infty$ Riemannian manifold (with boundary in general)
and $A_0$ an operator, both satisfying our general hypotheses. 
For any point $z\in {\cal M}$
(away from $\partial{\cal M}$) there is a geodesically
convex neighborhood of $z$,
${\cal N}_z$ (which does not intersect $\partial {\cal M}$)
such that, for any  pair 
$(x,y)\in {\cal N}_z$ 
\begin{eqnarray}
a_j(x,y) =  a_j(y,x)
\end{eqnarray}
for $j=0,1,2 \cdots$, where the heat-kernel coefficients are those given 
in} (\ref{vvm}) {\em and} (\ref{s}){\em .}\\

{\em Proof.} 
Fix any $z\in {\cal M}$ away from the boundary,  let  $\Omega'$ be an open set 
endowed with a coordinate frame $x^1,\cdots,x^D$ 
($\Omega'\cap \partial{\cal M}
=\emptyset$) and let $\Omega$ be a connected
 relatively-compact open neighborhood
of $z$ such that $\bar{\Omega}\subset \Omega'$. We can use the thesis
of {\bf Proposition 2.1} with the same notations employed there.
In particular, fix a  common open geodesically convex 
neighborhood ${\cal N}_z :={\cal N}_{zi_0}$ of $z$ 
 and its closure  given
in {\bf Proposition 2.1} and a sequence of analytic metrics 
$\{{\bf g}_n\}$
defined in a neighborhood of the compact set $\bar{\Omega}$
also given in {\bf Proposition 2.1}.
Since ${\cal N}_z$ and $\bar{{\cal N}}_{z}$ are geodesically convex
and, by {\bf Proposition 2.1},
 there is another similar open neighborhood 
${\cal N}_{zi}$, 
such that $\bar{{\cal N}}_{z} \subset {\cal N}_{zi}$, 
both coefficients $a_j(x,y)$
 and $a_j(y,x)$ are well-defined and smooth in $\bar{{\cal N}}_z
\times \bar{{\cal N}}_z$. 
For the moment, let us 
suppose also that 
$V$ is an analytic function of the considered coordinates.
Let us fix $x,y \in {\cal N}_z$, 
and consider the functionals of the metrics defined in $\bar{{\cal N}}_z$
\begin{eqnarray}
a_{jxy}[{\bf g}_n] &:=& (-1)^{j}\Delta^{-1/2}(x,y|{\bf g}_n) a_j(x,y|
{\bf g}_n) \:,
\label{xy}\\
a_{jyx}[{\bf g}_n] &:=& (-1)^{j}\Delta^{-1/2}(y,x|{\bf g}_n) 
a_j(y,x|{\bf g}_n)\:,
\label{yx}
\end{eqnarray}
$n=0, 1, \cdots \infty$, with ${\bf g}_\infty := {\bf g}$
and $\Delta := \Delta_{VVM}$.
If ${\bf g}_n$ is fixed, the coefficients above are smooth functions
on $\bar{\cal N}_z\times \bar{\cal N}_z$ which are also analytic in
$x$ and $y$.
Obviously, the symmetry of these functionals in $x,y$ would involve that of
the heat-kernel coefficients since the VVM determinant is symmetric. 
Since ${\bar{\cal N}_z}$ is totally normal,
it is possible to make explicit each $a_{jxy}$  and $a_{jxy}$ 
in terms of a sequence of integrals
computed along the unique geodesic between $x$ and $y$
which belongs completely to a normal neighborhood centered on 
$x$ (as well as  $y$) including the whole set $\bar{\cal N}_z$.
Moreover, since $\bar{\cal N}_z$ is geodesically convex with respect to all
the metrics, we can do it for all the metrics ${\bf g}_n$,
  the corresponding geodesics 
 depending on the particular metric one is considering.
Let us indicate 
the considered
geodesic starting from $y$ and reaching $y'\in \bar{\cal N}_z$ and 
computed with respect to the metric ${\bf g}_n$
by $\lambda \mapsto \gamma(\lambda,y',y|{\bf g}_n)$
(with $\lambda \in [0,1]$).
Employing (\ref{vvm}) and (\ref{s}), one finds, with
$A_{0x}[{\bf g}] =  -\nabla_{{\bf g}a}\nabla_{\bf g}^{a} + V$
\begin{eqnarray}
a_{0xy}[{\bf g}] &\equiv & 1\:,\nonumber \\
a_{1xy}[{\bf g }] &=& \int_0^1 d\lambda
[\Delta^{-1/2} A_0]_{\gamma(\lambda,x,y)}[{\bf g}] \Delta^{1/2}
 (\gamma(\lambda, x,y|{\bf g}),y|{\bf g}) \:,\nonumber \\
a_{2xy}[{\bf g}] &=&  \int_0^1d\lambda  \int_0^1 d\lambda' \lambda'
[\Delta^{-1/2} A_0]_{\gamma(\lambda',x,y)}[{\bf g}]
\Delta^{1/2}
(\gamma(\lambda',x,y|{\bf g}),y| {\bf g})
  \nonumber\\
& & \times 
[\Delta^{-1/2} A_0]_{\gamma(\lambda, \gamma(\lambda',x,y),y)}[{\bf g}]
\Delta^{1/2}(\gamma(\lambda, \gamma(\lambda',x,y|{\bf g}),y  |{\bf g}),y
|{\bf g})\:,
\nonumber\\
& & \cdots \nonumber\\
a_{jxy}[{\bf g}] &=& \int_0^1 d\lambda  \int_0^1 d\lambda_1 \lambda_1^1 \cdots
\int_0^1 d\lambda_{j-1} \lambda_{j-1}^{j-1} 
{\cal A}_{jxy}(\lambda,\lambda_1,\cdots
\lambda_{j-1}|{\bf g})
 \label{general} \:,   
\end{eqnarray}
where (omitting the explicit dependence on  the chosen metric
for sake of simplicity)
\begin{eqnarray}
{\cal A}_{jxy}(\lambda,\lambda_1,\cdots
\lambda_{j-1}) :=\nonumber 
\end{eqnarray}
\begin{eqnarray}
& & [\Delta^{-1/2} A_0]_{\gamma(\lambda_{j-1},x,y)}
\Delta^{1/2}(\gamma(\lambda_{j-1},x,y),y) \times \nonumber\\
& &
[\Delta^{-1/2} A_0]_{\gamma(\lambda_{j-2},\gamma(\lambda_{j-1},x,y),y)}
\Delta^{1/2}(\gamma(\lambda_{j-2},\gamma(\lambda_{j-1},x,y),y),y) \times 
\nonumber\\
& &\cdots\nonumber\\
& &  [\Delta^{-1/2} A_0]_{
\gamma(\lambda,\gamma(\lambda_1,\cdots 
\gamma(\lambda_{j-1},x,y),\cdots y),y)}
\Delta^{1/2} (\gamma(\lambda,\gamma(\lambda_1,\cdots 
\gamma(\lambda_{j-1},x,y),\cdots y),y),y)\:.
\end{eqnarray}
Notice that, fixing any metric,
$\Delta(x,y)$ and their derivatives are smooth functions of the 
derivatives of the function $\sigma(x,y)$ in the set $\bar{{\cal N}}_z
\times \bar{{\cal N}}_z$, therefore by item (c) of {\bf Proposition 2.1}
 and {\bf Lemma 2.2},
on the compact set $\bar{{\cal N}}_z\times \bar{{\cal N}}_z$,
one gets the 
uniform convergence with all of the derivatives of the
sequence of functions $\Delta^{\pm 1/2}(x,y|{\bf g}_n)$  to the function
$\Delta^{\pm 1/2}(x,y|{\bf g})$.
Moreover, from item (d) of {\bf Proposition 2.1}, taking into account 
that all the functions appearing in the integration above are computed
on the geodesics connecting $y$ with $x$ which, not depending on $n$,
 belong completely to  the compact  $\bar {{\cal N}}_z$, and
 using recurrently {\bf Lemma 2.2}
one gets\\  
(1)  for $j=0,1,2,\cdots$
\begin{eqnarray}
{\cal A}_{jxy}(\lambda,\lambda_1,\cdots
\lambda_{j-1}|{\bf g}_{n}) \rightarrow 
{\cal A}_{jxy}(\lambda,\lambda_1,\cdots
\lambda_{j-1}|{\bf g})\:, \label{limit}
\end{eqnarray}
 as $n \to +\infty$. This holds {\em uniformly} in
 $\lambda,\lambda_1, \cdots ,\lambda_{j-1} \in [0,1]$\\
and therefore,\\
(2) for any $j \in \N$, there is a constant $C_j$ such that
\begin{eqnarray}
|{\cal A}_{jxy}(\lambda,\lambda_1,\cdots
\lambda_{j-1}|{\bf g})| < C_j  \:\:\:\:\: \mbox{for}
\:\:\:\: n = 1,2, \cdots\:, \label{bounds}
\end{eqnarray}
 uniformly in
 $(\lambda,\lambda_1, \cdots ,\lambda_{j-1}) \in [0,1]^j$. 
Lebesgue's dominated convergence theorem assures that, for $n \to +\infty$, 
\begin{eqnarray}
a_{jxy}[{\bf g}_n] \rightarrow 
a_{jxy}[{\bf g}] \:.
\end{eqnarray}
The same result can be obtained considering the coefficients
$a_{jyx}[{\bf g}_n]$ and $a_{jyx}[{\bf g}]$. 
This allows one to conclude the proof noticing that,
\begin{eqnarray}
\left(a_{jxy}[{\bf g}_n] -a_{jyx}[{\bf g}_n]\right)
\to \left(a_{jxy}[{\bf g}] -a_{jyx}[{\bf g}]\right)
\end{eqnarray}
for $n \to +\infty$.
The left-hand side above vanishes because the metrics
${\bf g}_n$ are analytic in the open connected set ${\cal N}_z\times 
{\cal N}_z$ and thus the item 
(c) of {\bf Theorem 2.2} holds true.\\
If $V=V(x)$ is not analytic,
one can find a sequence
of positive analytic functions of the considered coordinates
in $\bar{\cal N}_y$, $\{ V_n\}$, such that this sequence converges 
uniformly to $V$ with all of its derivatives. 
This sequence can be obtained considering the convolutions
of $V$ and the flat-space heat kernel similarly to what we have done
in building up the sequence of the metrics ${\bf g}_n$ for proving 
{\bf Proposition 2.1} (see {\bf Appendix}).
Defining 
$A_{0x}[{\bf g}_n] :=  -\nabla_{{\bf g}_n}^{a}\nabla_{{\bf g}_n}^{a} + V_n$, 
and using the same arguments above,
 one can prove (\ref{limit}) and
(\ref{bounds}) once again and therefore 
gets the thesis. $\Box$\\

\noindent We have  a straightforward corollary based on the formulae
(\ref{unop}) and (\ref{unop'}).\\

\noindent  {\bf Corollary of Theorem 2.2.} {\em 
Let {\cal M} be a $D$-dimensional $C^\infty$ Riemannian manifold
and $A_0$ an operator both satisfying our general hypotheses.
For any point $z\in {\cal M}$
(away from $\partial{\cal M}$ 
if  $\partial{\cal M}$  is not empty) there is a geodesically
convex neighborhood ${\cal N}_y$  of $y$,
(which does not intersect $\partial {\cal M}$ 
if $\partial {\cal M}$  exists)
such that, for any pair $(x,y)\in {\cal N}_z$,
 up to the orders indicates in the summations below,
 the coefficients $u_j,v_j$ of the Hadamard parametrix
\begin{eqnarray}
H(x,y) &=& \sum_{j=0}^{D/2-2} 
\left(\frac{2}{\sigma(x,y)}\right)^{D/2-j-1} u_j(x,y) +
\sum_{j=0}^{2} v_j(x,y) \ln (\sigma(x,y)/2) 
\end{eqnarray}
for $D$ even (the summation appears for $D\geq 4$ only), and
\begin{eqnarray}
H(x,y) &=& \sum_{j=0}^{(D-5)/2} 
\left(\frac{2}{\sigma(x,y)}\right)^{D/2-j-1} u_j(x,y) + v(x,y) 
\sqrt{\frac{2\pi}{\sigma(x,y)}} \nonumber\\
& &+ w_0(x,y) \sqrt{2\pi \sigma(x,y)}
\end{eqnarray}
for $D$ odd (the summation  appears for $D\geq 5$ only), satisfy 
\begin{eqnarray}
u_j(x,y) &=&  u_j(y,x)\:, \\
v_j(x,y) &=&  v_j(y,x)\:. 
\end{eqnarray}
}

\noindent {\bf 2.2} {\em Conclusions and Outlooks.} We have proven the symmetry
of the heat-kernel/Hadamard  coefficients in the general Riemannian case. 
The Lorentzian case remains an open issue. However, we expect that one can
pass to the Lorentzian case from the Riemannian one by some analytic
continuation, if the manifold and the coefficient $V$ are analytic.
This should assure the symmetry of the considered coefficients
in the analytic Lorentzian case.
From that, the symmetry in the $C^\infty$ Lorentzian case is straightforward,
since the proof of {\bf Theorem 2.2} needs the validity of the symmetry
in the analytic case only, not depending on the signature of the metric.
Indeed, {\bf Proposition 2.1}, which is the kernel of the proof above,
holds true for any signature of the metric of the manifold (and some parts
of it can be generalized for non-metric affine connections).

\section*{Appendix: Proof of Proposition 2.1.}

 Several  simple lemmata are necessary. We do not report the proofs of those
lemmata for the sake of  brevity. These are based essentially 
on the Banach fixed-point theorem, the theorem of the inverse function 
and further simple considerations of elementary real 
analysis\footnote{A complete proof of the Lemmata contained in this appendix
can be found within the {\em first version} of the preprint
 gr-qc/9902034 (http://xxx.lanl.gov/abs/gr-qc/9902034v1).}. 
\\

\noindent {\bf Lemma A1.} {\em Let 
$f$ be a function of
$C^k([t_0-\Delta,t_0+\Delta]\times \bar{B}_R(y_0);\R^m)$, 
 where $k\in \{\infty,\omega\}$,
 $t_0$,
$\Delta>0$ and $R>0$ are real numbers and $B_R(y_0)$ indicates
the open ball of $\R^m$ centered in $y_0$ with radius $R$. Let us consider
the differential equation
\begin{eqnarray}
\frac{dY}{dt} =f(t,Y)\:\:\:\:\:\:\:\:\:\:\: 
Y\in C^{k}([t_0-\delta,t_0+\delta];\R^m) \:\:\:\:\:\: \mbox{for some }\:\:\:\
\delta>0, \delta\leq \Delta \label{equazione}
\end{eqnarray}
with initial condition
\begin{eqnarray}
Y(t_0) = \bar{y}_0 \:\:\:\:\:\:\: \bar{y}_0 \in \bar{B}_r(y_0),
\:\: r\:\: \mbox{fixed arbitrarily such that}\:\: 0<r<R \label{condizione}\:.
\end{eqnarray}}
(a) {\em A solution of Eq.} (\ref{equazione}) {\em with initial 
condition}
(\ref{condizione}) {\em exists and is unique in any set 
$[t_0-\delta, t_0+\delta]$, provided that
\begin{eqnarray}
0<\delta< Min\left(\Delta, \Delta'_r,\Delta''\right) \label{delta}\:,
\end{eqnarray}
where $\Delta'_r =[(R-r)/2]/\left\{
Sup\{||f(t,y)||\:\:|\:\: 
t\in [t_0-\Delta, t_0+\Delta] \: y \in \bar{B}_R(y_0)\}\right\}$ and\\
$\Delta'' = 1/\left\{Sup\{\sqrt{Tr \nabla f(t,y)^{T} \nabla f(t,y)}\:\:|\:\: 
t\in [t_0-\Delta, t_0+\Delta] \: y \in \bar{B}_R(y_0)\}\right\}$.}\\
(b) {\em It  satisfies $Y(t,\bar{y}_0) \in \bar{B}_R(y_0)$ for any 
$t \in [t_0-\delta,t_0+\delta]$ and
$\bar{y}_0\in \bar{B}_r(y_0)$.}\\
(c) {\em Moreover, the solution $(t,\bar{y}_0)\mapsto Y(t,\bar{y}_0)$
belongs to $C^{\infty}([t_0-\delta,t_0+\delta]\times \bar{B}_r(y_0);\R^m)$
and, in the case $k=\omega$, it is also 
analytic  
in the variable $t\in [t_0-\delta,t_0+\delta]$ and in the variable 
$\bar{y}_0 \in \bar{B}_r(y_0)$ (separately in general).}\\

\noindent {\bf Lemma A2}. {\em Let $\{f_n\}$ be a sequence of functions of
$C^\infty ([t_0-\Delta,t_0+\Delta]\times \bar{B}_R(y_0); \R^m)$, where the
 used notations are the same as those used in the {\bf Lemma A1}. Let us suppose 
also that, for any $p=0,1,2\cdots $ and for any multi-index $\alpha$, 
\begin{eqnarray}
D^\alpha_y \frac{\partial^p f_n}{\partial t^p} \to D_y^\alpha 
\frac{\partial^p f_\infty}{\partial t^p} \:\:\:\:\:
\mbox{uniformly on} \:\: [t_0-\Delta,t_0+\Delta]\times \bar{B}_R(y_0), 
\end{eqnarray}
$f_\infty$ being another function of
 $C^\infty ([t_0-\Delta,t_0+\Delta]\times \bar{B}_R(y_0); \R^m)$.\\
Let us indicate the solutions of the equation } (\ref{equazione}), {\em with
$f_n$in place of $f$ and initial condition} (\ref{condizione}), 
 {\em by $(t,\bar{y}_0) \mapsto
Y_n(t,\bar{y}_0)$ ($n= 0,1,2,\cdots , \infty$). Then,}
{\em for any $\delta>0$ satisfying} (\ref{delta}) {\em above with $f_\infty$
in place of $f$, and any $r>0$ with $r<R$:} \\
(a)  {\em There is a natural $N_\epsilon$ such that 
for $n>N_\epsilon$, $(t,\bar{y}_0) \mapsto  
Y_n(t,\bar{y}_0) $ is
 defined in $[t_0-\delta, t_0+\delta ]\times \bar{B}_r(y_0)$;}\\
(b)  {\em for any $p=0,1,2,\cdots$, 
\begin{eqnarray}
\frac{\partial^p Y_n}{\partial t^p} 
\to 
\frac{\partial^p Y_\infty}{\partial t^p} 
\:\:\:\:\:\:\: \mbox{uniformly in}\:\:   
[t_0-\delta,t_0+\delta]\times \bar{B}_r(y_0) \label{conv}\:.
\end{eqnarray}}\\

\noindent {\bf Lemma A3.} {\em With the same hypotheses of {\bf Lemma A2}
one gets also that,
for any $p=0,1,2,\cdots$ and for any multi-index $\alpha$, 
\begin{eqnarray}
D_{\bar{y}_0}^\alpha
\frac{\partial^p Y_n}{\partial t^p} 
\to D_{\bar{y}_0}^\alpha
\frac{\partial^p Y_\infty}{\partial t^p} 
\:\:\:\:\: \mbox{uniformly in}\:\:   
[t_0-\delta,t_0+\delta]\times \bar{B}_r(y_0) \label{conv'}\:.
\end{eqnarray}}\\

\noindent {\bf Lemma A4.} {\em Let $\{ f_n \}$ be a sequence of $\R^m$-valued 
functions defined in an open set $N\subset \R^m$ such that:}\\
(i) {\em   $f_n \in C^{k}(N;\R^m)$ for $n= 0, 1,\cdots$, where $k$ is fixed
in $\{\infty,\omega\}$ }.\\
(ii) {\em $f_n \to f_\infty  \in C^{k}(N;\R^m)$ uniformly 
in the set $N$ with all of their derivatives of any order.}\\
(iii) {\em There is a point $x_0\in N$ such that 
$det (\nabla f_\infty |_{x=x_0}) \neq 0$}.\\
 {\em Then, 
there exist two open neighborhoods of $x_0$ and $f_\infty(x_0)$
respectively, ${\cal U}_{x_0}$ and ${\cal V}_{f(x_0)}$,  a real $K>0$  and
a natural $N_0$ such that, for $n>N_0$ including $n=\infty$,}\\
(a) {\em all functions 
$f_n|_{{\cal U}_{x_0}} : {\cal U}_{x_0} \to f_n({\cal U}_{x_0})$
define diffeomorphisms, in particular, any $f_n({\cal U}_{x_0})$
is an open set. Moreover,
$|det (\nabla f_n|_{{\cal U}_{x_0}})| > K$.}\\
(b) 
\begin{eqnarray}
{\cal V}_{f_\infty(x_0)} \subset \bigcap_{n>N_0} f_n({\cal U}_{x_0}) 
\label{intersection}\:.
\end{eqnarray}
{\em where the intersection includes $n=\infty$.}\\
(c) {\em In the set ${\cal V}_{f_\infty(x_0)}$ and for $n\to \infty$,
$f_n^{-1} \to f^{-1}_\infty$ uniformly with all of their derivatives 
of any order. Moreover,
$0< |det \nabla 
(f^{-1}_n|_{{\cal V}_{f_\infty(x_0)}})| < \frac{1}{K}$}.
\\

\noindent {\bf Lemma A5.} {\em Let $K$ be a connected 
compact set of  $\R^m$ and $G, G_n: K \to  M(D,\R)$  continuous functions 
such that $G(x) =G(x)^T$ and  $G_n(x) =G_n(x)^T$ for any $x\in K$,
$n=0,1,2,\cdots$; $M(D,\R)$ denoting the  algebra  of  
real $D\times D$ matrices.
 Let us suppose that $G(x)$ is not singular for any $x\in K$ and, 
component by component, 
$G_n(x) \to G(x)$ uniformly in $x$, for $n\to \infty$.
Then, there is a natural $N_0$ such that,
for $n>N_0$ and not depending on $x\in K$,}
{\em all the matrices $G_n(x)$ are non-singular and
 $sign(G_n(x)) = sign( G(x))$,
  $sign(A)$ denoting the signature of the
    real symmetric matrix $A$.}\\

\noindent {\em Proof of }{\bf Proposition 2.1}.\\ 
Let us proceed with the proof of the item (a). Let $\Omega'$ be an
 open neighborhood of the point $z$ 
in the manifold ${\cal M}$, such that 
${\Omega'}\cap \partial{\cal M} =\emptyset$. Suppose also that
$\vec{x} \equiv (x^1,\cdots,x^D)$ are coordinates defined
in  $\Omega'$. Then, let $\Omega$ be a connected relatively-compact 
open neighborhood of $z$ such that $\bar{\Omega} \subset \Omega'$.\\
 Let ${\bf g}$ be the metric on ${\cal M}$
which can be either Lorentzian or Riemannian. Finally, let us define the
pure Euclidean-Laplacian heat kernel in $\R^D$,
\begin{eqnarray}
E(t,\vec{x},\vec{y}) :=
 \frac{e^{-||\vec{x} -\vec{y}||^2/4t}}{(4\pi t)^{D/2}}\:,
\label{E}
\end{eqnarray}
where $\vec{x},\vec{y} \in \R^D$ and $t\in ]0,+\infty[$.
From now on, we shall
identify the various subsets of $\Omega'$ with the corresponding subsets
of $\R^D$ through the given coordinate system.
Since the topology on $\Omega'$ is that of $\R^D$, one can find another 
connected 
relatively-compact open set $\Omega''$ such that $\bar{\Omega} \subset
\Omega''$ and $\bar{\Omega}''\subset \Omega'$. 
Let us consider the class of covariant second-order 
tensorial fields  defined in the given coordinate system on ${\Omega'}$, 
\begin{eqnarray}
g_{nab}(\vec{x}) := \int_{\R^D} d^D\vec{y}\: E(1/n,\vec{x},\vec{y})
g_{ab}(\vec{y}) \eta(\vec{y}) \:,\label{metrics}
\end{eqnarray}
where $d^D\vec{y}$ is the natural Lebesgue measure on $\R^D$ and
$\vec{x}\mapsto \eta(\vec{x})$ is a nonnegative $C^\infty$ function 
which takes the value  $1$ in $\bar{\Omega}$ and vanishes outside 
of $\bar{\Omega}''$. From the well-known properties of the Euclidean 
heat-kernel \cite{ch}
we have that, since $y\mapsto g_{ab}(\vec{y}) \eta(\vec{y})$ 
in (\ref{metrics}) 
is uniformly continuous in its domain
(as it is continuous in a compact set),
  $g_{nab}(\vec{x}) \to g_{ab}(\vec{x}) \eta(\vec{x})$
uniformly in  $\R^D$, as $n\to \infty$. In particular,
this holds in $\bar{\Omega}$ where $\eta(\vec{x}) = 1$. Therefore, for any 
point $\vec{x}\in \bar{\Omega}$, we have a sequence of symmetric matrices 
$G_n(\vec{x}) \equiv [g_{nab}(\vec{x})]$
which converges to the nonsingular symmetric matrix
$G(\vec{x}) \equiv [g_{ab}(\vec{x})]$ uniformly in $\vec{x}\in \bar{\Omega}$.
 By {\bf Lemma A5}, for $n>N_0$,
the matrices $G_n(\vec{x})$ define metrics in the 
tangent space at $\vec{x}$ with the same signature of $G(\vec{x})$.\\
The {\em uniform}
convergence in $\bar{\Omega}$
given above holds  also for the derivatives of any order
of the components $g_{nab}(\vec{x})$ and 
$g_{ab}(\vec{x})$. Indeed, from (\ref{metrics}), one has, passing 
the derivatives under the sign of integration (see the extended
discussion below) and then, using the integration by parts,
\begin{eqnarray}
D^\alpha_{\vec{x}}
g_{nab}(\vec{x}) 
&=& \int_{\R^D} d^D\vec{y}\: D^\alpha_{\vec{x}} E(1/n,\vec{x},\vec{y})
g_{ab}(\vec{y}) \eta(\vec{y})\nonumber \\
&=& \int_{\R^D} d^D\vec{y}\: (-1)^{|\alpha|}
(D^\alpha_{\vec{y}} E(1/n,\vec{x},\vec{y}))
g_{ab}(\vec{y}) \eta(\vec{y}) \nonumber \\
&=& \int_{\R^D} d^D\vec{y}\: 
 E(1/n,\vec{x},\vec{y})
D^\alpha_{\vec{y}}(g_{ab}(\vec{y})) \eta(\vec{y})  + 
G^{(\alpha)}(\vec{x}) \:.\label{Galpha} 
 \end{eqnarray}
The function $G^{(\alpha)}$ above is  
a sum of terms containing  derivatives of order $>0$ of the function 
$\eta$; omitting  overall constants, these terms 
are of the form ($|\gamma| >0$)
\begin{eqnarray}
 \int_{\R^D} d^D\vec{y}\:
 E(1/n,\vec{x},\vec{y})
D^\beta_{\vec{y}}
(g_{ab}(\vec{y})) D^\gamma_{\vec{y}} \eta(\vec{y}) \:.
\end{eqnarray}
Taking $\vec{x}\in \bar{\Omega}$ and $n\to \infty$ these terms 
converge to the functions $D^\beta_{\vec{x}}
(g_{ab}(\vec{x})) D^\gamma_{\vec{x}} \eta(\vec{x})$ which vanish
in $\bar{\Omega}$ since
$\eta$ is constant in  $\bar{\Omega}$  and $|\gamma|>0$.
 Therefore, dropping the term $G^{(\alpha)}$,
one has from (\ref{Galpha}), for $\vec{x} \in \bar{\Omega}$
and $n\to \infty$,
\begin{eqnarray}
D^\alpha_{\vec{x}}
g_{nab}(\vec{x}) \to D^\alpha_{\vec{x}}
g_{ab}(\vec{x}) \:,
\end{eqnarray}
uniformly in $\vec{x}$.\\
To conclude the proof of the item (a), 
let us prove that, fixing the indices $a,b$,
 the functions $\vec{x}\mapsto g_{ab}(\vec{x})$ are analytic functions of the 
coordinates $\vec{x}$ on the whole space $\R^D$.\\
From (\ref{metrics}) and the definition of the function $\eta$, we have
\begin{eqnarray}
g_{nab}(\vec{x}) := \int_{\bar{\Omega}''} d^D\vec{y}\: E(1/n,\vec{x},\vec{y})
g_{ab}(\vec{y})   \eta(\vec{y})  
\:.\label{metrics'}
\end{eqnarray}
Fixing the natural $n$ and
$\vec{y} \in {\bar{\Omega}}''$,  it is possible to continue the variable
$\vec{x}$ of the heat-kernel to complex values. It is obvious that,
fixing $n=1,2,\cdots$,
the obtained function $(\vec{\zeta},\vec{y}) \mapsto E(1/n,\vec{\zeta},
\vec{y})$
belongs to $C^\infty(\C^D\times\R^D)$ and, for any fixed $\vec{y}$,
 it is holomorphic in the variable $\vec{\zeta} \in \C$. Obviously, 
the derivatives of any order 
 in $Re \vec{\zeta}$ and $Im \vec{\zeta}$ of the integrand of (\ref{metrics})
are bounded in any compact set of the form
$\bar{\cal O}_{\vec{\zeta}_0}\times \bar{\Omega}''$, ${\cal O}_{\vec{\zeta}_0}$
being a relatively compact open neighborhood of $\vec{\zeta}_0\in \C^D $.
Therefore, Lebesgue's dominated convergence theorem implies that
the left hand side of (\ref{metrics'}) continued to complex values of 
$\vec{x} = \vec{\zeta}$ is smooth and one can pass the derivatives in (any 
component of) $Re \vec{\zeta}$ and $Im \vec{\zeta}$ 
under the sign of integration.
In this way one can check the validity of Cauchy-Riemann's conditions
 for the $\vec{y}$-integrated function of $\vec{\zeta}$
by the validity of the same conditions for the 
integrand function of $(\vec{\zeta},\vec{y})$. 
Therefore $\vec{\zeta} \mapsto g_{ab}(\vec{\zeta})$ is a 
complex-analytic function. Taking $Im \vec{\zeta} = 0$, one gets
the real-analyticity of the left-hand side of (\ref{metrics'}).

Let us pass to prove the item (b). Our strategy is the following: We
 shall define the 
exponential maps of each metric ${\bf g}_n$ around the point $z 
\equiv \vec{0}$ and thus, by using these exponential maps and
by shrinking the found  neighborhoods,
 we shall define   normal neighborhoods, totally normal
neighborhoods and geodesically convex neighborhoods of the metrics
${\bf g}_n$. Finally, we shall extract a class of 
(totally normal) geodesically convex
neighborhoods which are common to all metrics.\\
Let us fix $z\in \Omega$, define a normal coordinate system $\vec{y}$
with respect to the metric ${\bf g}_\infty := {\bf g}$ and
centered in $z \equiv \vec{0}$. This coordinate system is
defined in a normal neighborhood centered in ${z}$, 
${\cal N}_{{z}} = B_{\eta}(\vec{0})$, $\eta>0$, the open ball
above being defined in the normal coordinates with respect to the standard
$\R^D$ metric (obviously, 
this defines open sets with respect to the topology of the manifold).
Employing {\bf Lemma 2.2}, one sees that, in this system of coordinates
 one still has $g_{nab}(\vec{y})  \to g_{\infty ab}(\vec{y}) $ uniformly 
in $\vec{y}$, with all the derivatives. However, in general, the components
of the various metrics are not analytic functions of the coordinates, but
this is not important at this step.\\ 
In the considered coordinates, 
the {\em first order} geodesical equations read, 
for any metric ${\bf g}_n$ (including $n=\infty$),
\begin{eqnarray}
\frac{d y^{a}_n(t,\vec{y}_0,\vec{v}_0)}{dt} &=& 
v_n^{a}(t,\vec{y}_0,\vec{v}_0) \label{geo1}\:, \\
\frac{d v^{a}_n(t,\vec{y}_0,\vec{v}_0)}{dt} &=& -\Gamma_{nbc}^{a}(\vec{y}_n)
 v^{b}(t,\vec{y}_0,\vec{v}_0)  v^{c}(t,\vec{y}_0,\vec{v}_0) 
\label{geo2}\:.
\end{eqnarray}
(The sum over the 
repeated indices is understood).
Above, $\vec{y}_0$ and $\vec{v}_0$
 are, respectively, the initial position and the 
initial velocity evaluated at $t=0$. The latter, in components, defines a
vector in $T_{\vec{y}}({\cal M})$. 
From {\bf Lemma A1}, we know that the solution is unique
provided that $(t,(\vec{y}_0,\vec{v}_0)) \in [-\delta, \delta] \times 
\bar{B}_R((\vec{0},\vec{0}))$
 for some $\delta>0$ and $R>0$.
We can find a real $r>0$ such that 
$\bar{B}_r(\vec{0})\times\bar{B}_r(\vec{0})
\subset \bar{B}_R((\vec{0},\vec{0}))$
(in any case,  
$\bar{B}_r(\vec{0})\subset \Omega$ must hold). Obviously, the existence and uniqueness
of the solution holds true also replacing 
$\bar{B}_R((\vec{0},\vec{0}))$ by $\bar{B}_r(\vec{0})\times\bar{B}_r(\vec{0})$.
 Let us indicate  the geodesics given above in coordinates by $\gamma_n$.
From {\bf Lemma A2}
and {\bf Lemma A3}, we know that, in the considered common domain,  
employing coordinates $\vec{y}$
and for $n$ larger than some $N_0$, 
 $\gamma_n(t,\vec{y}_0,\vec{v}_0) \to
\gamma_\infty(t,\vec{y}_0,\vec{v}_0)$ with all the 
$t,\vec{y}_0,\vec{v}_0$ derivatives, uniformly in all these variables, where
$(t,\vec{y}_0,\vec{v}_0) \mapsto \gamma_\infty(t,\vec{y}_0,\vec{v}_0)$
is the geodesic associated to the target metric ${\bf g}_\infty ={\bf g}$.\\
For any fixed real $\alpha>0$,
(\ref{geo1}) and (\ref{geo2}) entail the identity 
\begin{eqnarray}
\gamma_n(\alpha t,\vec{y}_0,\vec{v}_0/\alpha)
= \gamma_n(t,\vec{y}_0,\vec{v}_0) \:,
\end{eqnarray}
for $n=0,1,\cdots,\infty$.
This means that, if $2 > \delta>0$, maintaining all properties concerning 
the uniform convergence and 
passing to the new variable $\lambda 
= (2/\delta) t$, we can work with geodesics
defined in the interval $\lambda\in [-2,2]$ provided $r$ is replaced by
$r' = (\delta/2) r < r$. Since there is no ambiguity we can  use the name
$r$ instead of  $r'$.
Therefore, from now on,   we suppose 
$\lambda \in [-2,2]$. This allows one to define the well-known
exponential maps for
$(\vec{y},\vec{v})\in \bar{B}_r(\vec{0})\times\bar{B}_r(\vec{0})$,
\begin{eqnarray}
(\vec{y},\vec{v}) \mapsto exp_{ny}(\vec{v}) := \gamma_n(1,\vec{y},\vec{v})
\label{expn}\:.
\end{eqnarray}
Once the exponential maps are defined in the common neighborhood above,
we can pass to study the totally normal neighborhoods.\\
To this end, 
let us consider the functions, defined in our coordinate system and in the
induced base in the tangent space,
\begin{eqnarray}
F_n :  \bar{B}_r(\vec{0})\times\bar{B}_r(\vec{0}) \to
{\cal M}\times{\cal M}: (\vec{y},\vec{v}) 
\mapsto (\vec{y}, exp_{ny}(\vec{v}))
\label{Fn}\:.
\end{eqnarray}
Notice that, in the considered domain, for $n\to \infty$,
the sequence of $F_n(\vec{y},\vec{v})$, like the sequences of their 
derivatives of any order in $(\vec{x},\vec{v})$,  converge to
$ F_\infty (\vec{y},\vec{v}) $ and corresponding derivatives of it,
uniformly in these variables.\\
Passing from the geodesic equations (\ref{geo1}),(\ref{geo2}) to the 
corresponding equations for the $\vec{y}$, $\vec{v}$-derivatives of 
the solutions, one can straightforwardly 
 prove that $det (\nabla F_\infty |_{(\vec{y},\vec{v})=
(\vec{0},\vec{0})}) = 1$. (Obviously, this property holds true for any $n$
and any point $\vec{y}$ in the considered domain.)
Therefore,  using {\bf Lemma A4} in the set
 $B_r(\vec{0})\times B_r(\vec{0})$, one gets  
that there is  a common neighborhood of $(\vec{0},\vec{0})$,
${\cal U}_{(\vec{0},\vec{0})} \subset B_r(\vec{0})\times B_r(\vec{0})$ where
all functions $F_n$, for $n>N'_0$, define diffeomorphisms.
Moreover there is an open neighborhood of
$F_\infty((\vec{0},\vec{0})) = (\vec{0},\vec{0})$,
${\cal V}_{(\vec{0},\vec{0})}$ such that
\begin{eqnarray}
{\cal V}_{(\vec{0},\vec{0})} 
\subset \bigcap_{n>N'_0} F_n({\cal U}_{(\vec{0},\vec{0})})\label{intrs}\:.
\end{eqnarray}
Without loss of generality, we can take ${\cal U}_{(\vec{0},\vec{0})}$
of the form ${\cal U}_{\vec{0}} \times B_\rho(\vec{0})$, $0<\rho<r$,
$\bar{{\cal U}}_{\vec{0}}$ being  an open neighborhood 
of $\vec{0}$.
Similarly, we can take 
${\cal V}_{(\vec{0},\vec{0})}$ of the form ${\cal V}_{\vec{0}}\times
{\cal V}_{\vec{0}}$. \\
In any open neighborhood of $(\vec{0},\vec{0})$,
 ${\cal V}_{(\vec{0},\vec{0})}$ satisfying (\ref{intrs}),
the inverse of the functions $F_n$  converges uniformly
with all the derivatives to the inverse of $F_\infty$, and 
$F_n^{-1}$ and $F_\infty^{-1}$ are diffeomorphisms. Therefore 
as $n\to\infty$, 
$exp_{n\vec{y}}^{-1}(\vec{y}') \to exp_{\infty \vec{y}}^{-1}(\vec{y}')$ 
uniformly in $(\vec{y},\vec{y}')\in {\cal V}_{\vec{0}}
\times{\cal V}_{\vec{0}}$.\\
This enable us  to prove that ${\cal V}_{\vec{0}}$ is
a totally normal neighborhood of $z\equiv \vec{0}$ 
for all the considered metrics.
Take $\vec{y} \in {\cal V}_{\vec{0}}$.
From the definition of $F_n$ and (\ref{intrs}), one has
\begin{eqnarray}
\{ \vec{y} \} \times {\cal V}_{\vec{0}}
 \subset F_n(\{ \vec{y}\} \times 
B_\rho(\vec{0})) \:, \nonumber
\end{eqnarray}
and therefore, $\vec{v} \mapsto exp_{n\vec{y}}(\vec{v})$
is a diffeomorphism in $B_\rho(\vec{0})$ and for any $n>N_0$ including
$n=\infty$,
\begin{eqnarray}
{\cal V}_{\vec{0}} \subset   exp_{n\vec{y}}(B_\rho(\vec{0}))
\:\:\:\:
\mbox{for any}\:\: \vec{y} \in {\cal V}_{\vec{0}}  \label{vrho}\:.
\end{eqnarray}
This means that ${\cal V}_{\vec{0}}$ is a totally normal neighborhood
of $z\equiv \vec{0}$
which is common for all metrics provided $n> N'_0$.\\
In this last step, we prove that it is possible to choose ${\cal V}_{\vec{0}}$
such that ${\cal V}_{\vec{0}}$ and $\bar{\cal V}_{\vec{0}}$ are common
geodesically convex neighborhoods of $z\equiv \vec{0}$
for all the metrics whenever
$n> N_0\geq N'_0$. Actually, we shall find a class of neighborhoods 
${\cal V}_{\vec{0}}$ defining a local base of the topology. 
Essentially we shall use the theory developed in the part 8 of  Chapter III
of \cite{kn}.\\
The set ${\cal V}_{\vec{0}}$ can be chosen as 
a ball $B_{\delta}(\vec{0})$. \\
Our thesis can be proven using the  following two results:

[1] There is an integer $N''_0>N'_0$ and a real $\rho'>0$,  
such that
for $n>N''_0$ (including $n=\infty$),
 the $D\times D$-matrix-valued  functions, given in components by
\begin{eqnarray}
A_{nab}(\vec{y}) := \left(\delta_{ab} -  \sum_{c} 
\Gamma_{nab}^{c}(\vec{y})y^c\right)
\label{quadratic}
\end{eqnarray}
are positive definite for $\vec{y} \in B_{\rho'}(\vec{0})$.\\
Above, the connection coefficients $\Gamma_{nab}^c$ are those corresponding
to the $n$-th metric {\em represented} in coordinates $\vec{y}$.

[2] Fixed an open ball $B_{\rho'}(\vec{0})$ with $\rho'>0$ sufficiently small,
it is possible to choose a natural 
$N_0>N''_0$, two reals  $\rho>0$ and $\delta'>0$  
 such that, for $n>N_0$ (including $n=\infty$),
using coordinates $\vec{y}$ in the domain as well as in the co-domain,
\begin{eqnarray}
exp_{n \vec{y}}(B_{\rho}(\vec{0})) \subset B_{\rho'}(\vec{0})
\:\:\:\:
\mbox{for any}\:\: \vec{y} \in B_{\delta'}(\vec{0})
 \label{cnd}\:.
\end{eqnarray}
Before we prove [1] and [2], we show that [1] and [2] entail
that for any real $\delta$ such that $0<\delta<\delta'$, one has
 ${\cal N}_z :=
{\cal V}_{\vec{0}} := B_\delta(\vec{0})$ and its closure 
is geodesically convex 
with respect to all metrics. 
The item (b) is therefore completely proved by posing, in
the considered coordinates,
${\cal N}_{zi} := B_\delta(\vec{0})$ with $i = -\cot (\pi \delta /\delta')$.
Notice that the open neighborhoods of $z$ given above define a local base of
the topology because one can define a Riemannian metric in a neighborhood of
$z$ such that the balls $B_\delta(\vec{0})$ are metric balls. Moreover, 
it is well known \cite{kn}
that the metric topology induced from any metric on a manifold coincides
with the topology of the manifold.\\
To this end, in our coordinate frame, let us 
indicate a geodesic of the $n$-th metric 
by $\lambda\mapsto \vec{y}_n(\lambda) $ 
and consider the function $\lambda\mapsto T_n(\lambda) = 
\sum_a (y^{a}(\lambda))^2$.
Suppose that such a geodesic is tangent to the boundary of a ball 
$\partial B_\epsilon(\vec{0})$ ($\epsilon>0$)
for $\lambda=0$. From the  geodesic equations, one gets
\begin{eqnarray}
\frac{d^2 T_n}{d\lambda^2}|_{\lambda=0} = 2\sum_{a,b} \left[\left
(\delta_{ab} -\sum_c 
\Gamma_{nab}^{c}(\vec{y}_n(0)) y_n^c(0)\right)
\frac{d^2y_n^{a}}{d\lambda^2}|_{\lambda=0} \frac{d^2y_n^{b}}
{d\lambda^2}|_{\lambda=0} 
\right] \:. \nonumber
\end{eqnarray}
Therefore [1] assures that, if $\epsilon<\rho'$, there is a neighborhood
of the tangent point where the geodesics lie outside $B_\epsilon(\vec{0})$
for $n>N''_0$ (including $n=\infty$).\\
Now we use [2] to conclude the proof. Let us choose $\rho$ in (\ref{vrho})
and $\delta'>0$,
such that (\ref{cnd}) is satisfied, $\rho'$ being that considered in [1].
We want to show that ${\cal V}_{\vec{0}} = B_\delta(\vec{0})$ 
and $\bar{{\cal V}}_{\vec{0}}$ 
are geodesically 
convex for any metrics ${\bf g}_n$, $n>N_0$ (including $n=\infty$)
if $0<\delta<\delta'$.
Let $\vec{y}_1$ and $\vec{y}_2$ be
 any couple of points of $B_\delta(\vec{0})$ (or $\bar{B}_\delta(\vec{0})$). 
Consider the $n$-geodesic $t\mapsto \vec{y}_n(\lambda)$, 
$\lambda\in [0,1]$, joining these points. We
want to show that it lies in $B_\delta(\vec{0})$ ($\bar{B}_\delta(\vec{0})$
respectively).
Suppose that this is not true. Then, there is at least one point 
 which does not  belong to$B_\delta(\vec{0})$ ($\bar{B}_\delta(\vec{0})$
 respectively).
In our hypotheses, the maximum of the function $T_{n}$
is attained for a 
value of the parameter 
$\lambda_{ne}\in ]0,1[$ since the extreme points of the geodesics 
belong to  $B_\delta(\vec{0})$ ($\bar{B}_\delta(\vec{0})$ respectively) 
and thus $T_n(0), T_n(1) < T_n(\lambda_{ne})$. Therefore, posing
$\vec{y}_{ne} := \vec{y}_n(\lambda_{ne})$ 
and $\rho_{ne} := T_{n}(\lambda_{ne})$, since $\lambda_{ne}$
is an {\em internal} point of $[0,1]$, 
 $dT_n/d\lambda|_{\lambda=\lambda_{ne}} =0$ must hold  and thus
the geodesic is tangent to $\partial 
B_{\rho_{ne}}(\vec{0})$ at
$\vec{y}_{ne}$ where $T_n$ reaches its maximum.
 Notice that,
because of [2], all the geodesics lie in $B_{\rho'}(\vec{0})$. Therefore,
due to [1], there is a neighborhood of $\vec{y}_{ne} = 
\vec{y}_n(\lambda_{ne})$ where the geodesics lie outside 
$B_{\rho_{ne}}(\vec{0})$. This is not possible.  This means that, not depending
on $n>N_0$, $B_\delta(\vec{0})$ and  $\bar{B}_\delta(\vec{0})$ 
are geodesically convex.
(Actually, the maximum of $T_n$ is attained at the extreme points where 
the geodesic is {\em not} tangent to the corresponding sphere,
and thus there is no {\em absurdum}.)\\
To conclude the proof of the item (b), let us prove [1] and [2] above.\\
The proof of [1] is quite simple. As we know, 
one has that, for a sufficiently small
ball centered in $\vec{0}$, the metrics ${\bf g}_n$, {\em represented in
coordinates} $\vec{y}$, converge uniformly to the metric ${\bf g}_\infty$
with all the derivatives. 
Therefore, in a sufficiently small ball centered in $\vec{0}$, 
$\Gamma_{nab}^{c}(\vec{y})$ must converge uniformly in $\vec{y}$
 to $\Gamma_{\infty ab}^{c}(\vec{y})$
for $n\to \infty$, and thus, from (\ref{quadratic}),
$||A_n -A_\infty||_\infty \to 0$.
Then notice that $\Gamma_{\infty ab}^c(\vec{0}) = 0$, because the coordinates
$\vec{y}$ are normal coordinates of the metric ${\bf g}_\infty$ centered on 
$\vec{0}$. Therefore, there is a sufficiently small ball centered in $\vec{0}$ 
where $A_\infty(\vec{y})$ 
is positive definite uniformly in $\vec{y}$, {\em i.e.}, there exists 
an $a>0$ 
such that $(u, A_\infty(\vec{y}) u) >a$ uniformly in $\vec{y}$
and  $u$ with $||u|| =1$. In fact, since $||u||=1$,
\begin{eqnarray}
|(u, (A_\infty(\vec{y}) - A_\infty(\vec{0})) u)| \leq 
||A_\infty(\vec{y}) - A_\infty(\vec{0}) || \:,
\end{eqnarray}
 and $||A_\infty(\vec{y}) - A_\infty(\vec{0}) || \to 0$
as $\vec{y} \to \vec{0}$. Since the bound above is uniform in $u$,
one has that, for any $\epsilon >0$,
there is a neighborhood of $\vec{y} =\vec{0}$ where,
\begin{eqnarray}
(u, A_\infty(\vec{y}) u) > (u, A_\infty(\vec{0}) u) -\epsilon = 1-\epsilon\:,
\end{eqnarray}
uniformly in $u$.
 Taking $\epsilon>0$ such that $1-\epsilon = a>0$, $a$ is the requested 
positive lower bound of $(u, A_\infty(\vec{y}) u)$.
Now, we can repeat the same procedure considering the norm $||\:||_\infty$
and different values of $n$ in the found neighborhood. One has
\begin{eqnarray}
|(u, (A_n(\vec{y}) - A_\infty(\vec{y})) u)| \leq 
||A_\infty - A_n ||_\infty \:.
\end{eqnarray}
Since $||A_n - A_\infty||_\infty \to 0$ as $n\to 0$, we get that, for any 
$\epsilon>0$, there is a $N''_0$ such that, for $n>N''_0$,
\begin{eqnarray}
(u, A_n(\vec{y}) u) > (u, A_\infty(\vec{y}) u) -\epsilon > a-\epsilon\:,
\end{eqnarray}
 uniformly in
$u$ and $\vec{y}$.
Taking $0<\epsilon<a$, one has proventhe thesis.\\
Let us prove [2]. The case $n = \infty$ is trivial since the function
$(\vec{y},\vec{v}) \mapsto exp_{\infty \vec{y}}(\vec{v})$
 is continuous
and $exp_{\infty \vec{0}}(\vec{0}) = \vec{0}$.
We also know that, for sufficiently small $\rho,\delta>0$, in 
 $B_{\delta'}(\vec{0})\times B_{\rho}(\vec{0})$,
the sequence of functions
 $(\vec{y},\vec{v}) \mapsto exp_{n \vec{y}}(\vec{v})$ converges
to the function
$(\vec{y},\vec{v}) \mapsto exp_{\infty \vec{y}}(\vec{v})$
uniformly in $(\vec{y},\vec{v})$ as $n\to \infty$. 
In $B_{\delta'}(\vec{0})\times B_{\rho}(\vec{0})$, one has
\begin{eqnarray}
||exp_{n \vec{y}}(\vec{v}) - exp_{\infty \vec{0}}(\vec{0}) ||
&\leq& ||exp_{n \vec{y}}(\vec{v}) - exp_{\infty \vec{y}}(\vec{v}) || +
||exp_{\infty \vec{y}}(\vec{v}) - exp_{\infty \vec{0}}(\vec{0}) || \nonumber\:.
\end{eqnarray}
Moreover, fixing $\rho'>0$, one can take  $\epsilon>0$
such that $0<\rho'-\epsilon < \rho'/2$, and find a pair $\rho,\delta'>0$
such that for  $(\vec{y},\vec{v})\in 
B_{\delta'}(\vec{0})\times B_{\rho}(\vec{0})$
\begin{eqnarray}
||exp_{\infty \vec{0}}(\vec{0}) - exp_{\infty \vec{y}}(\vec{v}) ||
< \rho'-\epsilon\:,\nonumber
\end{eqnarray}
and  a natural $N_0$ such that, on the same  ball and for 
$n>N_0$,
\begin{eqnarray}
||exp_{n \vec{y}}(\vec{v}) - exp_{\infty \vec{y}}(\vec{v}) ||
< \rho'-\epsilon\:.\nonumber
\end{eqnarray}
Therefore, in $B_{\delta'}(\vec{0})\times B_{\rho}(\vec{0})$, one has
\begin{eqnarray}
||exp_{n \vec{y}}(\vec{v}) - exp_{\infty \vec{0}}(\vec{0}) ||
&\leq& (\rho' -\epsilon)+ (\rho' -\epsilon) < \rho'\nonumber\:.
\end{eqnarray}
This is the thesis.

The items (c) and (d) are trivially proven by noticing that, 
as a consequence of the analogous property of the diffeomorphisms  $F_n$
defined above, 
in the
 normal coordinates $\vec{y}$ and thus in any other coordinate system
around $z \equiv \vec{0}$ which covers any  set ${\em N}_{zi}$, the 
diffeomorphisms $(\vec{y},\vec{y}')\mapsto exp_{n\vec{y}}^{-1}(\vec{y'})$
converge to the function $(\vec{y},\vec{y}')\mapsto 
exp_{\infty\vec{x}}^{-1}(\vec{y}'))$ uniformly with all the derivatives.
Similarly, the geodesics $(\lambda,\vec{y},\vec{v}) \mapsto
\gamma_n(\lambda,\vec{y},\vec{v}) $ 
converge uniformly in all arguments
jointly to the geodesic
$(\lambda,\vec{y},\vec{v}) \mapsto
\gamma_\infty(\lambda,\vec{y},\vec{v}) $ with all the derivatives
for $(\vec{y},\vec{v}) \in B_r(\vec{0})\times B_r(\vec{0})$.
Employing our procedure to define the neighborhoods ${\cal N}_{zi}$ 
given above,
it is possible to shrink them, maintaining  all the relevant properties,
in such a way that $\bar{\cal N}_{zi} \subset 
 B_r(\vec{0})$, $i\in \R$. By consequence, as $n\to \infty$,
 the sequence of functions
\begin{eqnarray}
(\lambda,\vec{y},\vec{y}'  )\mapsto \gamma_n(\lambda,y,y') = 
\vec{y}_n(\lambda,y, exp_{n\vec{y}}^{-1}(\vec{y}'))
\end{eqnarray}
 defined on  any set $[0,1]\times \bar{\cal N}_{zi}\times \bar{\cal N}_{zi}$,
and the sequence of  functions 
\begin{eqnarray}
\sigma_n(y,y') = {\bf g}_n(\vec{y})(exp_{n\vec{y}}^{-1}(\vec{y}'),
exp_{n\vec{y}}^{-1}(\vec{y}'))\:,
\end{eqnarray}
defined on any set $\bar{\cal N}_{zi}\times \bar{\cal N}_{zi}$,
 converge uniformly in all the variables jointly, to the 
corresponding functions computed with respect to the metric
${\bf g}_\infty ={\bf g}$. Making a recurrent use of {\bf Lemma 2.2},
 this result can be proven also concerning the 
derivatives of any order in all variables and in any  coordinate
system. $\Box$\\

\noindent {\em Acknowledgement.} 
I am particularly indebted to  A. Cassa for his constant
assistance in solving mathematical problems related to this work
and for his  numerous and always illuminating technical  suggestions.
It is a pleasure to thank
 I. G. Avramidi, E. Ballico,  S. Delladio, 
 F. Serra Cassano, A. Tognoli and R. M. Wald for helpful discussions
and G. Esposito, B. S. Kay and D. Klemm for valuable suggestions.
This work  has been financially supported
by a Postdoctoral Research Fellowship of the Department of Mathematics
of the University of Trento.

 \end{document}